\documentclass{article}

\usepackage{amsmath,amsthm,amssymb}
\usepackage{graphicx}
\usepackage{wrapfig}
\usepackage{comment}

\usepackage{tikz}

\usetikzlibrary{shapes, arrows}
\usetikzlibrary{automata}

\theoremstyle{remark}
\newtheorem{example}{Example}

\tikzstyle{mol} = [fill,circle,inner sep=1pt]

\DeclareMathOperator*{\argmin}{arg\,min}

\title{A general architecture of oritatami systems for simulating arbitrary finite automata}
\author{Yo-Sub Han, Hwee Kim, Yusei Masuda, and Shinnosuke Seki}

\begin{document}

\maketitle

\begin{abstract}
In this paper, we propose an architecture of oritatami systems with which one can simulate an arbitrary nondeterministic finite automaton (NFA) in a unified manner. 
The oritatami system is known to be Turing-universal but the simulation available so far requires 542 bead types and $O(t^4 \log^2 t)$ steps in order to simulate $t$ steps of a Turing machine. 
The architecture we propose employs only 329 bead types and requires just $O(t |Q|^4 |\Sigma|^2)$ steps to simulate an NFA over an input alphabet $\Sigma$ with a state set $Q$ working on a word of length $t$. 
\end{abstract}

	\section{Introduction}

\begin{figure}[htb]
\centering
\includegraphics[width=\linewidth]{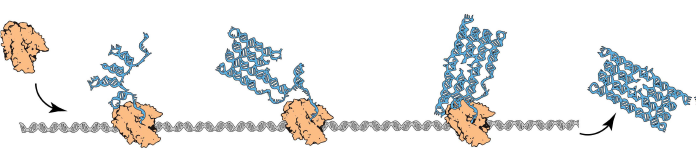}
\caption{RNA origami, a novel self-assembly technology by cotranscriptional folding \cite{GearyRothemundAndersen2014}. 
RNA polymerase (orange complex) attaches to an artificial template DNA sequence (gray spiral) and synthesizes the complementary RNA sequence (blue sequence), which folds into a rectangular tile while being synthesized. 
}
\label{fig:rna_origami}
\end{figure}

\textit{Transcription} (Figure~\ref{fig:rna_origami}) is a process in which from a template DNA sequence, its complementary RNA sequence is synthesized by an \textit{RNA polymerase} letter by letter. 
The product RNA sequence (\textit{transcript}) is folding upon itself into a structure while being synthesized. 
This phenomenon called \textit{cotranscriptional folding} has proven programmable by Geary, Rothemund, and Andersen in \cite{GearyRothemundAndersen2014}, in which they programmed an RNA rectangular tile as a template DNA sequence in the sense that the transcript synthesized from this template folds cotranscriptionally into that specific RNA tile highly probably \textit{in vitro}. 
As cotranscriptional folding has turned out to play significant computational roles in organisms (see, e.g., \cite{WaStYuLiLu2016}), a next step is to program computation in cotranscriptional folding. 

\textit{Oritatami} is a mathematical model proposed by Geary et al.~\cite{GeMeScSe2016} to understand computational aspects of cotranscriptional folding. 
This model has recently enabled them to prove that cotranscriptional folding is actually Turing universal \cite{GeMeScSe2018}. 
Their Turing-universal oritatami system $\Xi_{\rm TU}$ adopts a periodic transcript\footnote{A periodic transcript is likely to be able to be transcribed from a circular DNA \cite{GearyAndersen2014}.} whose period consists of functional units called \textit{modules}. 
Some of these modules do computation by folding into different shapes, which resembles somehow computation by cotranscriptional folding in nature \cite{WaStYuLiLu2016}. 
Being thus motivated, the study of cotranscriptional folding of shapes in oritatami was initiated by Masuda, Seki, and Ubukata in \cite{MasudaSekiUbukata2018} and extended independently by Domaine et al.~\cite{DHOPRSST2018} as well as by Han and Kim \cite{HanKim2018} further. 
In \cite{MasudaSekiUbukata2018}, an arbitrary finite portion of the Heighway dragon fractal was folded by an oritatami system $\Xi_H$. 
The Heighway dragon can be described as an automatic sequence \cite{AlloucheShallit2003}, that is, as a sequence producible by a deterministic finite automaton with output (DFAO) in an algorithmic manner. 
The system $\Xi_{\rm HD}$ involves a module that simulates a 4-state DFAO $A_{\rm HD}$ for the Heighway dragon. 
The Turing-universal system was not embedded into $\Xi_{\rm HD}$ in place for this module primarily because it may fold into different shapes even on inputs of the same length and secondly because it employs unnecessarily many 542 types of abstract molecules (\textit{bead}) along with an intricate network of interactions (rule set) among them. 
Their implementation of the DFAO module however relies on the cycle-freeness of $A_H$ too heavily to be generalized for other DFAs; let alone for nondeterministic FAs (NFAs). 

In this paper, we propose an architecture of oritatami system that allows for simulating an arbitrary NFA using 329 bead types. 
In order to run an NFA over an alphabet $\Sigma$ with a state set $Q$ on an input of length $t$, it takes $O(t|Q|^4|\Sigma|^2)$ steps (stabilization of this number of beads). 
In contrast, the system $\Xi_{\rm TU}$ requires $O(t^4 \log^2 t)$ steps to simulate $t$ steps of a Turing machine. 
A novel feature of technical interest is that all the four modules of the architecture share a common interface (\eqref{format:state} in Section~\ref{sect:architecture}).

	\section{Preliminaries}

Let $B$ be a set of types of abstract molecules, or \textit{beads}, and $B^*$ be the set of finite sequences of beads including the empty sequence $\lambda$. 
A bead of type $b \in B$ is called a $b$-bead. 
Let $w = b_1 b_2 \cdots b_n \in B^*$ be a sequence of length $n$ for some integer $n$ and bead types $b_1, \ldots, b_n \in B$. 
For $i, j$ with $1 \le i, j \le n$, let $w[i..j]$ refer to the subsequence $b_i b_{i+1} \cdots b_j$ of $w$; we simplify $w[i..i]$ as $w[i]$. 

The oritatami system folds its transcript, which is a sequence of beads, over the triangular grid graph $\mathbb{T} = (V, E)$ cotranscriptionally based on hydrogen-bond-based interactions (\textit{h-interaction} for short) which the system allows for between beads of particular types placed at the unit distance. 
When beads form an h-interaction, we say informally they are \textit{bound}. 
The $i$-th point of a directed path $P = p_1 p_2 \cdots p_n$ in $\mathbb{T}$ is referred to as $P[i]$, that is, $P[i] = p_i$. 
A \textit{(finite) conformation} $C$ is a triple $(P, w, H)$ of a directed path $P$ in $\mathbb{T}$, $w \in B^*$ of the same length as $P$, and a set of h-interactions $H \subseteq \bigl\{ \{i, j\} \bigm| 1 \le i, i{+}2 \le j, \{P[i], P[j]\} \in E \bigr\}$. 
This is to be interpreted as the sequence $w$ being folded in such a manner that its $i$-th bead is placed at the $i$-th point of the path $P$ and the $i$-th and $j$-th beads are bound iff $\{i, j\} \in H$. 
A symmetric relation $R \subseteq B \times B$ called \textit{rule set} governs which types of two beads can form an h-interaction between. 
An h-interaction $\{i, j\} \in H$ is \textit{valid with respect to $R$}, or \textit{$R$-valid}, if $(w[i], w[j]) \in R$. 
A conformation is $R$-valid if all of its h-interactions are $R$-valid. 
For $\alpha \ge 1$, a conformation is \textit{of arity $\alpha$} if it contains a bead that forms $\alpha$ h-interactions and none of its beads forms more. 
By $\mathcal{C}_{\le \alpha}$, we denote the set of all conformations of arity at most $\alpha$. 

An oritatami system grows conformations by elongating them according to its own rule set $R$. 
Given an $R$-valid finite conformation $C_1 = (P, w, H)$, we say that another conformation $C_2$ is its \textit{elongation by a bead of type $b \in B$}, written as $C_1 \xrightarrow{R}_b C_2$, if $C_2 = (Pp, wb, H \cup H')$ for some point $p$ not along the path $P$ and possibly-empty set of h-interactions $H' \subseteq \bigl\{ \{i, |w|+1\} \bigm| 1 \le i < |w|, \{P[i], p\} \in E, (w[i], b) \in R\bigr\}$. 
Observe that $C_2$ is also $R$-valid. 
This operation is recursively extended to the elongation by a finite sequence of beads as: 
$C \xrightarrow{R}_\lambda^* C$ for any conformation $C$; and 
$C_1 \xrightarrow{R}_{wb}^* C_2$ for conformations $C_1, C_2$, a finite sequence of beads $w \in \Sigma^*$, and a bead $b \in \Sigma$ if there is a conformation $C'$ such that $C_1 \xrightarrow{R}_w^* C'$ and $C' \xrightarrow{R}_b C_2$. 

A finite \textit{oritatami system} is a tuple $\Xi = (R, \alpha, \delta, \sigma, w)$, where $R$ is a rule set, $\alpha$ is an arity, $\delta \ge 1$ is a parameter called \textit{delay}, $\sigma$ is an $R$-valid initial conformation of arity at most $\alpha$ called \textit{seed}, upon which its finite transcript $w \in B^*$ is to be folded by stabilizing beads of $w$ one at a time so as to minimize energy collaboratively with its succeeding $\delta{-}1$ nascent beads. 
The \textit{energy} of a conformation $C = (P, w, H)$, denoted by $\Delta G(C)$, is defined to be ${-}|H|$; that is, more h-interactions make a conformation more stable. 
The set $\mathcal{F}(\Xi)$ of conformations \textit{foldable} by this system is recursively defined as: 
the seed $\sigma$ is in $\mathcal{F}(\Xi)$; and 
provided that an elongation $C_i$ of $\sigma$ by the prefix $w[1..i]$ be foldable (i.e., $C_0 = \sigma$), its further elongation $C_{i+1}$ by the next bead $w[i{+}1]$ is foldable if 
\begin{equation}\label{eq:cotranscriptional_folding}
C_{i+1} \in \argmin_{
\substack{
C \in \mathcal{C}_{\le \alpha} s.t. \\
C_i \xrightarrow{R}_{w[i{+}1]}C \\
}
}
\min \Big\{ \Delta G(C') \Bigm| 
C \xrightarrow{R}^*_{w[i{+}2...i{+}k]}C', k\le \delta, C' \in \mathcal{C}_{\le \alpha}
\Big\}.
\end{equation}
We say that the bead $w[i{+}1]$ and the h-interactions it forms are \textit{stabilized} (not nascent any more) according to $C_{i+1}$. 
Note that an arity-$\alpha$ oritatami system cannot fold any conformation of arity larger than $\alpha$. 
The system $\Xi$ is \textit{deterministic} if for all $i \ge 0$, there exists at most one $C_{i+1}$ that satisfies \eqref{eq:cotranscriptional_folding}. 
An oritatami system is \textit{cyclic} if its transcript admits a period shorter than the half of itself. 

\begin{figure}[tb]
\centering
\scalebox{0.7}{\begin{tikzpicture}

\foreach \x in {0, 4, 8, 13} {
\draw[thick, red] (\x, 0) node[mol] {} node[above] {\large ${\tt 585}$}
-- ++(300:1) node[mol] {} node[left] {${\tt 586}$} 
-- ++(240:1) node[mol] {} node[below] {\large ${\tt 587}$}
-- ++(0:1) node[mol] {} node[below] {\large ${\tt 588}$}
-- ++(60:1) node[mol] {}  node[left] {${\tt 589}$}
-- ++(120:1) node[mol] {} node[above] {\large ${\tt 590}$}
;
\draw[thick, dashed,red] (\x, 0) -- ++(0:1) -- ++(240:1);
}

\draw[cyan, -latex] (1, 0) 
-- ++(120:1) node[mol] {} node[above] {\large ${\tt 579}$} 
-- ++(180:1) node[mol] {} node[above] {${\tt 580}$}
-- ++(240:1) node[mol] {} node[above] {\large ${\tt 581}$};
\draw[cyan, -latex] (1, 0) 
-- ++(60:1) node[mol] {} node[above] {\large ${\tt 579}$} 
-- ++(0:1) node[mol] {} node[above] {${\tt 580}$}
-- ++(300:1) node[mol] {} node[below] {\large ${\tt 581}$};
\draw[ultra thick, cyan, -latex] (1, 0) 
-- ++(0:1) node[mol] {} node[above] {\Large ${\tt 579}$} 
-- ++(300:1) node[mol] {} node[left] {${\tt 580}$}
-- ++(240:1) node[mol] {} node[below] {\Large ${\tt 581}$};

\draw[dashed, cyan] (0,0)++(300:2) -- ++(0:1) ++(60:1) -- ++(180:1);

\draw (3,0)++(300:1) node {\Large $\Rightarrow$};
\draw (7,0)++(300:1) node {\Large $\Rightarrow$};
\draw (12,0)++(300:1) node {\Large $\Rightarrow$};

\draw[thick, -latex] (5, 0) -- ++(0:1) node[mol] {} node[above] {\large ${\tt 579}$};

\draw[cyan, -latex] (6, 0) 
-- ++(300:1) node[mol] {} node[left] {${\tt 580}$}
-- ++(240:1) node[mol] {} node[below] {\large ${\tt 581}$} -- ++(0:1) node[mol] {} node[below] {\large ${\tt 582}$};
\draw[cyan, -latex] (5,0)++(300:2)++(240:0.3) -- ++(240:0.7) node[mol] {} node[below] {\large ${\tt 582}$};
\draw[cyan, -latex] (5,0)++(300:2.3) -- ++(300:0.7) node[mol] {} node[below] {\large ${\tt 582}$};
\draw[dashed, cyan] (4,0)++(300:2) -- ++(0:1)++(60:1) -- ++(180:1);

\draw[thick, -latex] (9, 0) -- ++(0:1) node[mol] {} node[above] {\large ${\tt 579}$}
-- ++(300:1) node[mol] {} node[left] {${\tt 580}$}
;

\draw[cyan, -latex] (10,0)++(300:1) -- ++(240:1) node[mol] {} node[below] {\large ${\tt 581}$}
-- ++(240:1) node[mol] {} node[below] {\large ${\tt 582}$}
-- ++(180:1) node[mol] {} node[below] {${\tt 583}$}
;
\draw[cyan, -latex] (9,0)++(300:2) 
-- ++(0:1) node[mol] {} node[below] {\large ${\tt 582}$} 
-- ++(60:1) node[mol] {} node[left] {${\tt 583}$};
\draw[cyan, -latex] (10,0)++(300:1) 
-- ++(60:1) node[mol] {} node[right] {\large ${\tt 581}$}
-- ++(120:1) node[mol] {} node[above] {\large ${\tt 582}$}
-- ++(180:1) node[mol] {} node[above] {${\tt 583}$}
;

\draw[dashed] (9,0)++(300:1) -- ++(0:1);
\draw[cyan,dashed] (8,0)++(300:2) -- ++(0:1);

\draw[thick, -latex] (14, 0) 
-- ++(0:1) node[mol] {} node[above] {\large ${\tt 579}$}
-- ++(300:1) node[mol] {} node[left] {${\tt 580}$}
-- ++(240:1) node[mol] {} node[below] {\large ${\tt 581}$}
;
\draw[dashed] (13,0)++(300:2) -- ++(0:1)++(60:1) -- ++(180:1);
\draw[cyan, -latex] (14,0)++(300:2) 
-- ++(0:1) node[mol] {} node[below] {\large ${\tt 582}$}
-- ++(60:1) node[mol] {} node[left] {${\tt 583}$}
-- ++(120:1) node[mol] {} node[above] {\large ${\tt 584}$}
;
\draw[cyan,dashed] (15,0) -- ++(0:1);

\end{tikzpicture}}

\caption{Growth (folding) of a spacer of glider shape (g-spacr). 
The rule set $R$ to fold this is $\{({\tt 579}, {\tt 584})$, $({\tt 580}, {\tt 589}), ({\tt 581}, {\tt 588}),  ({\tt 582}, {\tt 587}), ({\tt 583}, {\tt 586}), ({\tt 585}, {\tt 590}), ({\tt 586}, {\tt 590})\}$.}
\label{fig:glider}
\end{figure}
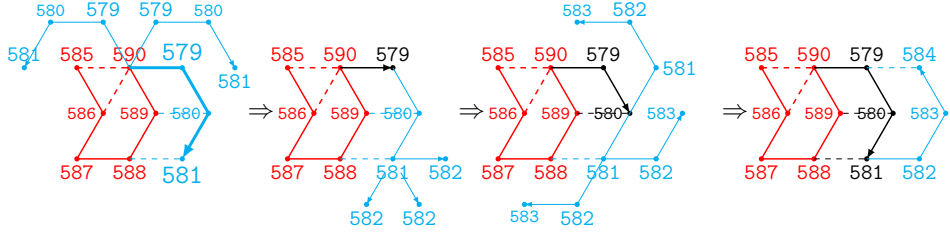

\begin{example}[g-spacer]
	Let us provide an example of deterministic oritatami system that folds into a \textit{glider} motif, which will be used as a component called \textit{g-spacer} in Section~\ref{sect:architecture}. 
	Consider a delay-3 oritatami system whose transcript $w$ is a repetition of ${\tt 579}{-}{\tt 580}{-}\cdots{-}{\tt 590}$ and rule set $R$ is as captioned in Figure~\ref{fig:glider}. 
	Its seed, colored in red, can be elongated by the first three beads $w[1..3] = {\tt 579}{-} {\tt 580}{-}{\tt 581}$ in various ways, only three of which are shown in Figure~\ref{fig:glider} (left). 
	The rule set $R$ allows $w[1]$ to be bound to ${\tt 584}$, $w[2]$ to ${\tt 589}$, and $w[3]$ to ${\tt 588}$, but {\tt 584}-bead is not around.  
	In order for both $w[2]$ and $w[3]$ to be thus bound, the nascent fragment $w[1..3]$ must be folded as bolded in Figure~\ref{fig:glider} (left). 
	According to this most stable elongation, the bead $w[1] = {\tt 579}$ is stabilized to the east of the previous ${\tt 580}$-bead. 
	Then $w[4] = {\tt 582}$ is transcribed. 
	It is capable of binding to a ${\tt 587}$-bead but no such bead is reachable, and hence, this newly-transcribed bead cannot override the ``bolded'' decision. 
	Therefore, $w[2]$ is also stabilized according to this decision along with its bond with the {\tt 589}-bead. 
	The next bead $w[5] = {\tt 583}$ cannot override the decision, either, and hence, $w[3]$ is stabilized along with its bond with the {\tt 588}-bead as shown in Figure~\ref{fig:glider} (right). 
\end{example}

	\section{Architecture}
	\label{sect:architecture}

We shall propose an architecture of a nondeterministic cyclic oritatami system $\Xi$ that simulates at delay 3 an NFA $A = (Q, \Sigma, q_0, Acc, f)$, where $Q$ is a finite set of states, $\Sigma$ is an alphabet, $q_0 \in Q$ is the initial state, $Acc \subseteq Q$ is a set of accepting states, and $f: Q \times \Sigma \to 2^Q$ is a nondeterministic transition function. 
What the architecture actually simulates is rather its modification $A_{\$} = (Q \cup \{q_{Acc}\}, \Sigma \cup \{\$\}, q_0, \{q_{Acc}\}, f \cup f_\$)$, where $f_\$: (Q \cup \{q_{Acc}\}) \times (\Sigma \cup \{\$\}) \to 2^{Q \cup \{q_{Acc}\}}$ is defined over $\Sigma$ exactly same as $f$, and moreover, $f_\$(q, \$) = \{q_{Acc}\}$ for all $q \in Acc$. 
Note that $w \in L(A)$ iff $w\$ \in L(A_{\$})$. 
For the sake of upcoming arguments, we regard the transition function $f \cup f_\$$ rather as a set of transitions $\{f_1, f_2, \ldots, f_n\}$, where $f_k$ is a triple $(o_k, a_k, t_k)$, meaning that reading $a_k$ in the state $o_k$ (origin) causes a transition to $t_k$ (target). 
Note that $n = O(|Q|^2|\Sigma|)$. 

The architecture assumes that each state $q$ is uniquely assigned with an $n$-bit binary sequence, whose $i$-th bit from most significant bit (MSB) is referred to as $q[i]$.
It also assumes a unique $m$-bit binary sequence for each letter $a \in \Sigma$, whose $\ell$-th bit from MSB is referred to as $a[\ell]$, where $m = \lceil \log |\Sigma| \rceil$. 

\subsection{Overview}

\begin{figure}[tb]
\centering
\includegraphics[width=\linewidth]{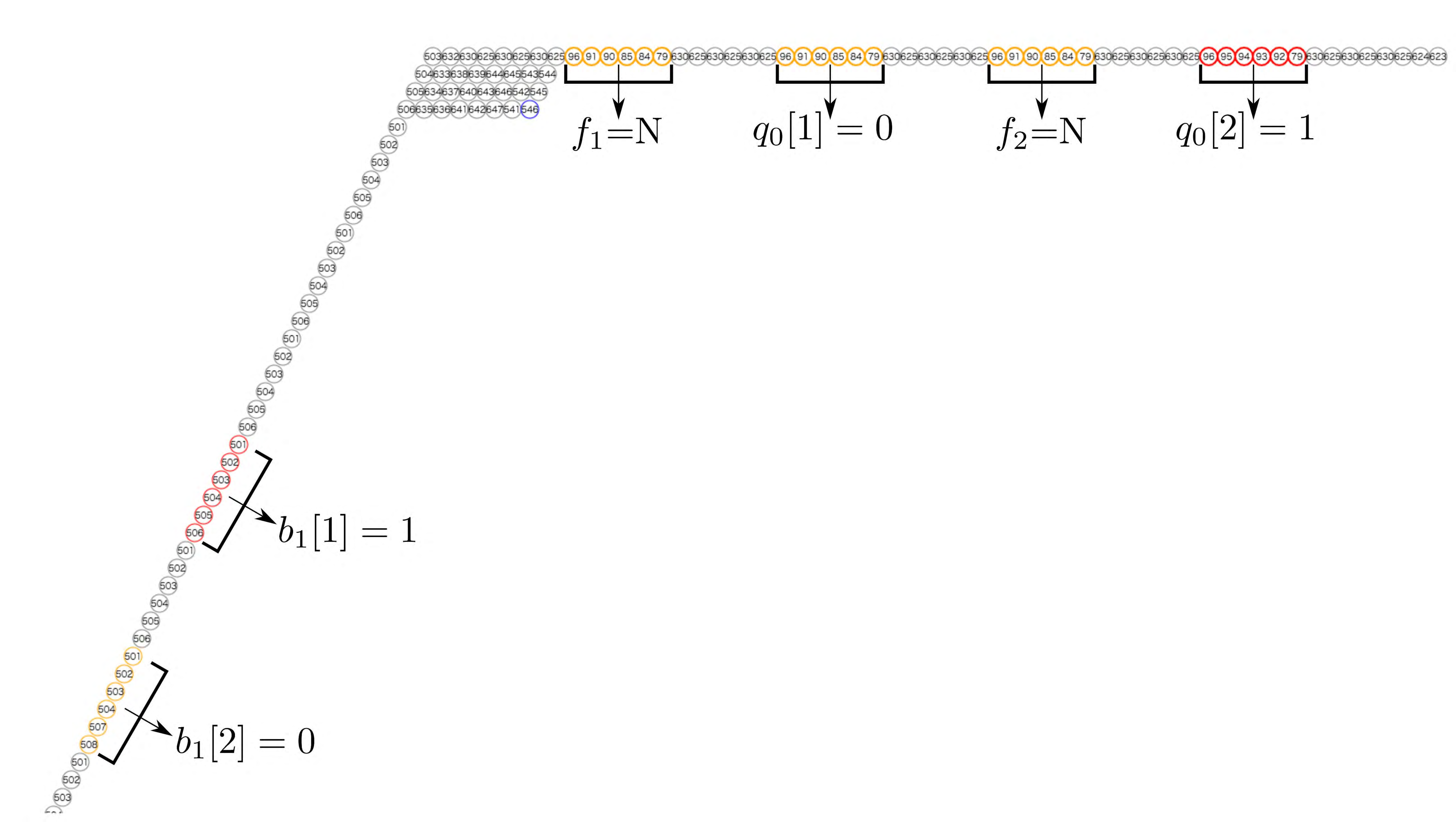}
\caption{Encoding of the initial state $q_0 = 01$ and the first letter $b_1 = 10$ of an input word on the seed of $\Gamma$-shape.}
\label{fig:seed}
\end{figure}

\paragraph{Seed.}
The seed of $\Xi$ is of $\Gamma$ shape as shown in Figure~\ref{fig:seed}. 
Below its horizontal arm is encoded the initial state $q_0$ in the following format: 
\begin{equation}\label{format:state}
{\tt \mbox{$\bigodot_{k=1}^n$} \bigl(
\mbox{$x_{f_k}$} \to (630 \to 625 \to)^3 \mbox{$z_{q_0[k]}$} \to (630 \to 625 \to)^3 \bigr) 624 \to 623, 
}
\end{equation}
where 
$z_0 = {\tt 96 \to 91 \to 90 \to 85 \to}$ ${\tt 84 \to 79}$, 
$z_1 = {\tt 96 \to 95 \to 94}$ ${\tt \to 93 \to 92 \to 79}$, and 
for some bead types $b, c \in B$, the arrow $b \to c \ \mbox{(resp. $\nearrow, \nwarrow, \gets, \swarrow, \searrow$)}$ implies that a $c$-bead is located to the eastern (resp. north-eastern, north-western, western, south-western, and south-eastern) neighbor of a $b$-bead. 
Note that the seed and Module 4, which we shall explain soon, initialize all the variables $f_1, \ldots, f_n$ to $N$ by having a sequence $x_N$ of bead types be exposed to the corresponding positions $x_{f_1}, \cdots, x_{f_n}$, where $x_N = z_0$ while $x_Y = z_1$, which is not used here but shall be used later. 
An input word $u = b_1b_2 \cdots$ is encoded on the right side of its vertical arm as: 
\begin{equation}\label{format:input_letter}
\bigodot_{j=1}^{|u|} \biggl( (y_{sp} \swarrow)^{2|f|-1} \mbox{$\bigodot_{\ell=1}^m$} \bigl(\mbox{$y_{b_i[\ell]}$} \swarrow \mbox{$y_{sp}$} \swarrow \bigr) (y_{sp} \swarrow)^{2+2|f|} \biggr), 
\end{equation}
where $y_{sp} = y_1 = {\tt 501 \swarrow 502 \swarrow 503 \swarrow 504 \swarrow 505 \swarrow 506}$ and $y_0 = {\tt 501 \swarrow 502}$ ${\tt \swarrow 503 \swarrow 504 \swarrow 507 \swarrow 508}$. 

The first period of the transcript of $\Xi$ starts folding at the top left corner of the seed, or more precisely, as to succeed its last bead {\tt 540} circled in blue in Figure~\ref{fig:seed}. 
It folds into a parallelogram macroscopically by folding in a zigzag manner microscopically (zig ($\hookrightarrow$) and zag ($\hookleftarrow$) are both of height 3) while reading the current (initial) state $q_0$ from above and the first letter $b_1$ from left, and outputs one of the states in $f(q_0, b_1)$, say $q_1$, nondeterministically below in the format \eqref{format:state}. 
All the states in $f(q_0, b_1)$ are chosen equally probably. 
The folding ends at the bottom left corner of the parallelogram. 
The next period likewise reads the state $q_1$ and next letter $b_2$, and outputs a state in $f(q_1, b_2)$ nondeterministically below the parallelogram it has folded. 
Generally speaking, for $i \ge 2$, the $i$-th period simulates a nondeterministic transition from the state $q_{i-1}$, output by the previous period, on the letter $b_i$. 

\paragraph{Modules.}
One period of the transcript is semantically factorized into four functional subsequences called modules. 
All these modules fold into a parallelogram of width $\Theta(n)$ and of respective height $6 \times 2n$, $6 \times 2m$, $6 \times 2$, and $6 \times 2n$ (recall one zigzag is of height 6); that is, the first module makes $2n$ zigzags, for example. 
These parallelograms pile down one after another. 
One period thus results in a parallelogram of width and height both $\Theta(n)$.  
Their roles are as follows: 
\begin{description}
\item[Module 1] extracts all the transitions that originate at the current state; 
\item[Module 2] extracts all the transitions that read the current letter among those chosen by Module 1;
\item[Module 3] nondeterministically chooses one state among those chosen by Module 2, if any, or halts the system otherwise; 
\item[Module 4] outputs the chosen state downward. 
\end{description}
In this way, these modules filter candidates for the next transition, and importantly, through a common interface, which is the format \eqref{format:state}. 

	\subsection{Implementation}

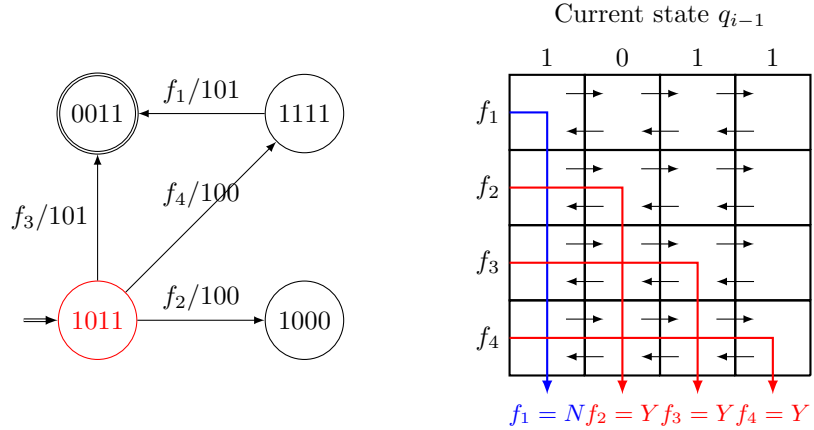
\begin{figure}[tb]
\centering
\begin{minipage}{0.45\linewidth}
\scalebox{1}{
\begin{tikzpicture}[>=latex, node distance=2.75cm,  initial text=, bend angle=15]
		 \tikzstyle{every initial by arrow} = [->, double];

		 \node [initial, state,red] (A)                         {$1011$};
		 \node [state, accepting]                     (B) [above of = A]  {$0011$};
		 \node [state]                     (C) [right of = A] {$1000$};
		 \node [state]                     (D) [right of = B] {$1111$};

 		\path [->] (A) edge [right] node [above]              {$f_2 /100$}                 (C)
         		         edge [right] node [left]             {$f_3 /101$}               (B)
		         edge [right] node [above]             {$f_4 /100$}               (D)
         			   (D) edge [left] node [above]             {$f_1 /101$}                 (B);
\end{tikzpicture}}
\end{minipage}
\begin{minipage}{0.05\linewidth}
\ \\
\end{minipage}
\begin{minipage}{0.45\linewidth}
\scalebox{1}{\begin{tikzpicture}

\foreach \x in {5}{
\foreach \y in {1,2,3,4}{
\draw[thick] (\x, \y)-- ++(90:1)-- ++(180:1)-- ++(270:1)-- ++(0:1);
\draw[thick] (\x+1, \y)-- ++(90:1)-- ++(180:1)-- ++(270:1)-- ++(0:1);
\draw[thick] (\x+2, \y)-- ++(90:1)-- ++(180:1)-- ++(270:1)-- ++(0:1);
\draw[thick] (\x+3, \y)-- ++(90:1)-- ++(180:1)-- ++(270:1)-- ++(0:1);
}}

\foreach \x in {5,6,7}{
\foreach \y in {1,2,3,4}{
\draw[-latex] (\x, \y+0.75)++(180:0.25)-- ++(0:0.5);
\draw[-latex] (\x, \y+0.25)++(0:0.25)-- ++(180:0.5); 
}}

\foreach \x in {5}{
\foreach \y in {4}{
\draw (\x,\y+1)++ (180:0.5) node [above] {$1$};
\draw (\x+1,\y+1)++ (180:0.5) node [above] {$0$};
\draw (\x+2,\y+1)++ (180:0.5) node [above] {$1$};
\draw (\x+3,\y+1)++ (180:0.5) node [above] {$1$};
\draw (\x+1,\y+1.5)++ (0:0) node [above] {Current state $q_{i-1}$};

\draw (\x-1,\y)++ (90:0.5) node [left] {$f_1$};
\draw (\x-1,\y-1)++ (90:0.5) node [left] {$f_2$};
\draw (\x-1,\y-2)++ (90:0.5) node [left] {$f_3$ };
\draw (\x-1,\y-3)++ (90:0.5) node [left] {$f_4$};

\draw[thick,-latex,blue] (\x-1,\y+0.5)-- ++(0:0.5)-- ++(270:3.75) node [below] {\small{$f_1 =N$}};
\draw[thick, -latex,red] (\x-1,\y-0.5)-- ++(0:1.5)-- ++(270:2.75) node [below] {\small{$f_2 =Y$}};
\draw[thick, -latex,red] (\x-1,\y-1.5)-- ++(0:2.5)-- ++(270:1.75) node [below] {\small{$f_3 =Y$}};
\draw[thick, -latex,red] (\x-1,\y-2.5)--++ (0:3.5)-- ++(270:0.75) node [below] {\small{$f_4 =Y$}};
}}
\end{tikzpicture}}
\end{minipage}

\caption{Example run of the proposed architecture.
(Left) A 4-state FA with 4 transitions $f_1, f_2, f_3, f_4$ to be simulated, which is obtained from a 3-state FA with the 2 transitions $f_2$ and $f_4$ in the way explained in the text, that is, by adding a new accepting sink state 0011 and transitions $f_1, f_3$ on the special letter \$, encoded as 101.
(Right) Outline of the flow of 1-bit information of whether each of the transitions originates from the current state 1011 or not through Module~1. 
}
\label{fig:example_FA_module1}
\end{figure}

\begin{figure}[tb]
\centering

\scalebox{1}{\begin{tikzpicture}
\foreach \x in {5}{
\foreach \y in {2,3,4}{
\draw[thick] (\x, \y)-- ++(90:1)-- ++(180:1)-- ++(270:1)-- ++(0:1);
\draw[thick] (\x+1, \y)-- ++(90:1)-- ++(180:1)-- ++(270:1)-- ++(0:1);
\draw[thick] (\x+2, \y)-- ++(90:1)-- ++(180:1)-- ++(270:1)-- ++(0:1);
\draw[thick] (\x+3, \y)-- ++(90:1)-- ++(180:1)-- ++(270:1)-- ++(0:1);
}}

\foreach \x in {5,6,7}{
\foreach \y in {2,3,4}{
\draw[-latex] (\x, \y+0.75)++(180:0.25)-- ++(0:0.5);
\draw[-latex] (\x, \y+0.25)++(0:0.25)-- ++(180:0.5); 
}}

\foreach \x in {5}{
\foreach \y in {4}{
\draw (\x,\y+1)++ (180:0.5) node [above] {\small{$f_1=N$}};
\draw (\x+1,\y+1)++ (180:0.5) node [above] {\small{$f_2=Y$}};
\draw (\x+2,\y+1)++ (180:0.5) node [above] {\small{$f_3=Y$}};
\draw (\x+3,\y+1)++ (180:0.5) node [above] {\small{$f_4=Y$}};

\draw (\x-1.25,\y)++ (90:0.5) node [left] {$1$};
\draw[thick, ->>] (\x-1.3,\y+0.5)-- ++(0:0.3);
\draw (\x-1.25,\y-1)++ (90:0.5) node [left] {$0$};
\draw[thick, ->>] (\x-1.3,\y-0.5)-- ++(0:0.3);
\draw (\x-1.25,\y-2)++ (90:0.5) node [left] {$0$};
\draw[thick, ->>] (\x-1.3,\y-1.5)-- ++(0:0.3);
\draw (\x-2,\y-1)++ (90:0.5) node [left] {\rotatebox{-90}{$b_i =100$}};

\draw[thick,-latex,blue] (\x-0.5,\y+1)-- ++(270:3) node [below] {\small{$f_1=N$}};
\draw[thick, -latex,red] (\x+0.5,\y+1)-- ++(270:3) node [below] {\small{$f_2=Y$}};
\draw[thick, red] (\x+1.5,\y+1)-- ++(270:2);
\draw[thick, -latex,blue] (\x+1.5,\y-1)-- ++(270:1) node [below] {\small{$f_3=N$}};
\draw[thick, -latex,red] (\x+2.5,\y+1)-- ++(270:3) node [below] {\small{$f_4=Y$}};

}}
\end{tikzpicture}}

\caption{Example run of the oritatami system constructed according to the proposed architecture in order to simulate the FA in Figure~\ref{fig:example_FA_module1}.
Here Module~2 filters transitions $f_2, f_3, f_4$ chosen by Module 1 further depending on whether each of them reads the letter 100 or not; thus $f_3$ is out. 
}
\label{fig:example_module2}
\end{figure}
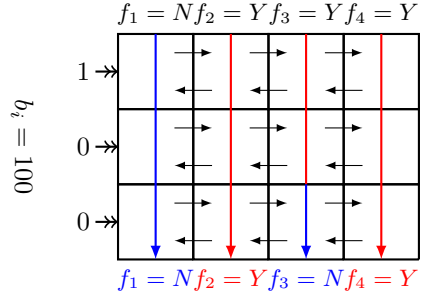

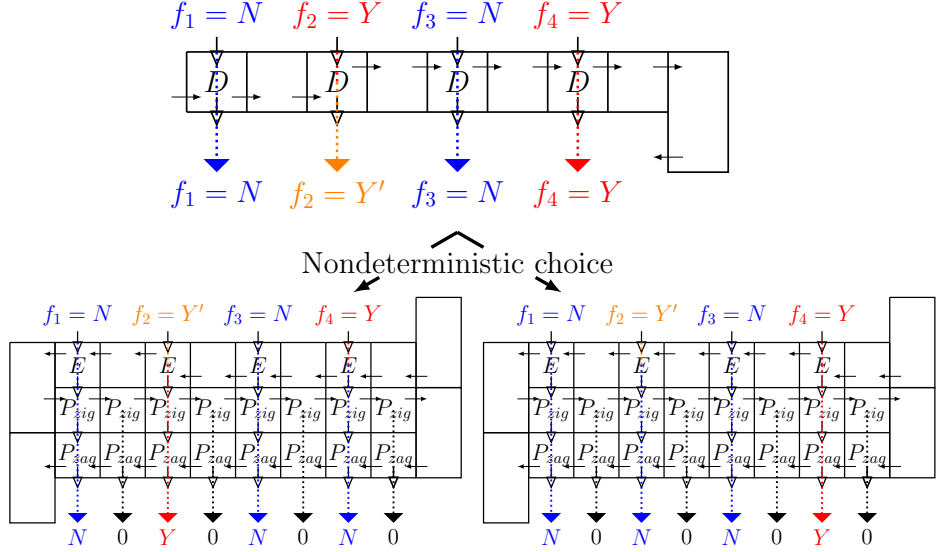
\begin{figure}[tb]
\centering
\scalebox{0.8}{\begin{tikzpicture}
\foreach \x in {0}{
\foreach \y in {10,11}{
\draw[thick] (\x, \y)-- ++(0:8);
}}
\foreach \x in {0,1,2,3,4,5,6,7,8}{
\foreach \y in {10}{
\draw[thick] (\x, \y)-- ++(90:1);
}}

\foreach \x in {1}{
\foreach \y in {11.25}{
\draw[thick, blue] (\x-0.5,\y) node [above] {\Large{$f_1=N$}};
\draw[thick, red] (\x+1.5,\y) node [above] {\Large{$f_2=Y$}};
\draw[thick, blue] (\x+3.5,\y) node [above] {\Large{$f_3=N$}};
\draw[thick, red] (\x+5.5,\y) node [above] {\Large{$f_4=Y$}};
}}

\foreach \x in {0,2,4,6}{
\foreach \y in {10,11}{
\draw[thick, -open triangle 45] (\x+0.5,\y) ++ (90:0.25) -- ++ (270:0.5);
}}

\foreach \x in {0}{
\foreach \y in {10}{
\draw[-latex] (\x, \y+0.25)++(180:0.25)-- ++(0:0.50); 
\draw (\x+0.5, \y+0.5) node {\Large{$D$}};
\draw[-latex] (\x+1, \y+0.25)++(180:0.25)-- ++(0:0.50); 
\draw[-latex] (\x+2, \y+0.25)++(180:0.25)-- ++(0:0.50); 
\draw (\x+2.5, \y+0.5) node {\Large{$D$}};
\draw[-latex] (\x+3, \y+0.75)++(180:0.25)-- ++(0:0.50); 
\draw[-latex] (\x+4, \y+0.75)++(180:0.25)-- ++(0:0.50); 
\draw (\x+4.5, \y+0.5) node {\Large{$D$}};
\draw[-latex] (\x+5, \y+0.75)++(180:0.25)-- ++(0:0.50); 
\draw[-latex] (\x+6, \y+0.75)++(180:0.25)-- ++(0:0.50); 
\draw (\x+6.5, \y+0.5) node {\Large{$D$}};
\draw[-latex] (\x+7, \y+0.75)++(180:0.25)-- ++(0:0.50); 
\draw[-latex] (\x+8, \y+0.75)++(180:0.25)-- ++(0:0.50); 
\draw[thick] (\x+8, \y+1)-- ++(0:1)-- ++(270:2)-- ++(180:1)-- ++(90:1);
\draw[-latex] (\x+8, \y-0.75)++(0:0.25)-- ++(180:0.50); 
}}

\foreach \x in {0}{
\foreach \y in {11}{
\draw[-triangle 90, blue, dotted, very thick] (\x+0.5, \y) -- ++(270:2) node [below] {\Large{$f_1 = N$}};
\draw[red, dotted, very thick] (\x+2.5, \y) -- ++(270:0.5);
\draw[-triangle 90, orange, dotted, very thick] (\x+2.5, \y-0.5) -- ++(270:1.5) node [below] {\Large{$f_2 = Y^\prime$}};
\draw[-triangle 90, blue, dotted, very thick] (\x+4.5, \y) -- ++(270:2) node [below] {\Large{$f_3 = N$}};
\draw[-triangle 90, red, dotted, very thick] (\x+6.5, \y) -- ++(270:2) node [below] {\Large{$f_4 = Y$}};

\draw[-latex, ultra thick] (\x+4.5, \y-3) -- ++(210:0.5)++(210:1)--++(210:0.5);
\draw[-latex, ultra thick] (\x+4.5, \y-3) -- ++(330:0.5)++(330:1)--++(330:0.5);
\node at (\x+4.5, \y-3.5) {\Large Nondeterministic choice};
}}

\end{tikzpicture}}

\centering
\begin{minipage}{0.45\linewidth}
\scalebox{0.6}{
\begin{tikzpicture}
\foreach \x in {0}{
\foreach \y in {8,9,10,11}{
\draw[thick] (\x, \y)-- ++(0:8);
}}
\foreach \x in {0,1,2,3,4,5,6,7,8}{
\foreach \y in {8}{
\draw[thick] (\x, \y)-- ++(90:3);
}}

\foreach \x in {1}{
\foreach \y in {11.25}{
\draw[thick, blue] (\x-0.5,\y) node [above] {\Large{$f_1=N$}};
\draw[thick, orange] (\x+1.5,\y) node [above] {\Large{$f_2=Y^\prime$}};
\draw[thick, blue] (\x+3.5,\y) node [above] {\Large{$f_3=N$}};
\draw[thick, red] (\x+5.5,\y) node [above] {\Large{$f_4=Y$}};
}}

\foreach \x in {0,2,4,6}{
\foreach \y in {8,9,10,11}{
\draw[thick, -open triangle 45] (\x+0.5,\y) ++ (90:0.25) -- ++ (270:0.5);
}}

\foreach \x in {1,3,7,7}{
\foreach \y in {8}{
\draw[thick, -open triangle 45] (\x+0.5,\y) ++ (90:0.25) -- ++ (270:0.5);
}}

\foreach \x in {0}{
\foreach \y in {10}{
\draw[-latex] (\x, \y+0.75)++(0:0.25)-- ++(180:0.5); 
\draw (\x+0.5, \y+0.5) node {\Large{$E$}};
\draw[-latex] (\x+1, \y+0.75)++(0:0.25)-- ++(180:0.5); 
\draw[-latex] (\x+2, \y+0.75)++(0:0.25)-- ++(180:0.5); 
\draw (\x+2.5, \y+0.5) node {\Large{$E$}};
\draw[-latex] (\x+3, \y+0.25)++(0:0.25)-- ++(180:0.5); 
\draw[-latex] (\x+4, \y+0.25)++(0:0.25)-- ++(180:0.5); 
\draw (\x+4.5, \y+0.5) node {\Large{$E$}};
\draw[-latex] (\x+5, \y+0.25)++(0:0.25)-- ++(180:0.5); 
\draw[-latex] (\x+6, \y+0.25)++(0:0.25)-- ++(180:0.5); 
\draw (\x+6.5, \y+0.5) node {\Large{$E$}};
\draw[-latex] (\x+7, \y+0.25)++(0:0.25)-- ++(180:0.5); 
\draw[-latex] (\x+8, \y+0.25)++(0:0.25)-- ++(180:0.5); 
\draw[thick] (\x+8, \y+2)-- ++(0:1)-- ++(270:2)-- ++(180:1)-- ++(90:2);
}}

\foreach \x in {0}{
\foreach \y in {9}{
\draw[thick] (\x-1, \y+2)-- ++(0:1)-- ++(270:2)-- ++(180:1)-- ++(90:2);
\draw[-latex] (\x, \y+0.75)++(180:0.25)-- ++(0:0.5); 
\draw (\x+0.5, \y+0.5) node {\Large{$P_{zig}$}};
\draw[-latex] (\x+1, \y+0.75)++(180:0.25)-- ++(0:0.5); 
\draw (\x+1.5, \y+0.5) node {\Large{$P_{zig}$}};
\draw[-latex] (\x+2, \y+0.75)++(180:0.25)-- ++(0:0.5); 
\draw (\x+2.5, \y+0.5) node {\Large{$P_{zig}$}};
\draw[-latex] (\x+3, \y+0.75)++(180:0.25)-- ++(0:0.5); 
\draw (\x+3.5, \y+0.5) node {\Large{$P_{zig}$}};
\draw[-latex] (\x+4, \y+0.75)++(180:0.25)-- ++(0:0.5); 
\draw (\x+4.5, \y+0.5) node {\Large{$P_{zig}$}};
\draw[-latex] (\x+5, \y+0.75)++(180:0.25)-- ++(0:0.5); 
\draw (\x+5.5, \y+0.5) node {\Large{$P_{zig}$}};
\draw[-latex] (\x+6, \y+0.75)++(180:0.25)-- ++(0:0.5); 
\draw (\x+6.5, \y+0.5) node {\Large{$P_{zig}$}};
\draw[-latex] (\x+7, \y+0.75)++(180:0.25)-- ++(0:0.5); 
\draw (\x+7.5, \y+0.5) node {\Large{$P_{zig}$}};
\draw[-latex] (\x+8, \y+0.75)++(180:0.25)-- ++(0:0.5); 
}}

\foreach \x in {0}{
\foreach \y in {8}{
\draw[thick] (\x-1, \y+1)-- ++(0:1)-- ++(270:2)-- ++(180:1)-- ++(90:2);
\draw[-latex] (\x, \y+0.25)++(0:0.25)-- ++(180:0.5); 
\draw (\x+0.5, \y+0.5) node {\Large{$P_{zag}$}};
\draw[-latex] (\x+1, \y+0.25)++(0:0.25)-- ++(180:0.5); 
\draw (\x+1.5, \y+0.5) node {\Large{$P_{zag}$}};
\draw[-latex] (\x+2, \y+0.25)++(0:0.25)-- ++(180:0.5); 
\draw (\x+2.5, \y+0.5) node {\Large{$P_{zag}$}};
\draw[-latex] (\x+3, \y+0.25)++(0:0.25)-- ++(180:0.5); 
\draw (\x+3.5, \y+0.5) node {\Large{$P_{zag}$}};
\draw[-latex] (\x+4, \y+0.25)++(0:0.25)-- ++(180:0.5); 
\draw (\x+4.5, \y+0.5) node {\Large{$P_{zag}$}};
\draw[-latex] (\x+5, \y+0.25)++(0:0.25)-- ++(180:0.5); 
\draw (\x+5.5, \y+0.5) node {\Large{$P_{zag}$}};
\draw[-latex] (\x+6, \y+0.25)++(0:0.25)-- ++(180:0.5); 
\draw (\x+6.5, \y+0.5) node {\Large{$P_{zag}$}};
\draw[-latex] (\x+7, \y+0.25)++(0:0.25)-- ++(180:0.5); 
\draw (\x+7.5, \y+0.5) node {\Large{$P_{zag}$}};
\draw[-latex] (\x+8, \y+0.25)++(0:0.25)-- ++(180:0.5); 
\draw[thick] (\x+8, \y+2)-- ++(0:1)-- ++(270:2)-- ++(180:1)-- ++(90:2);
}}

\foreach \x in {0}{
\foreach \y in {11}{
\draw[-triangle 90, blue, dotted, very thick] (\x+0.5, \y) -- ++(270:4) node [below] {\Large{$N$}};
\draw[-triangle 90, dotted, very thick] (\x+1.5, \y-1.5) -- ++(270:2.5) node [below] {\Large{$0$}};
\draw[orange, dotted, very thick] (\x+2.5, \y) -- ++(270:0.5);
\draw[-triangle 90, red, dotted, very thick] (\x+2.5, \y-0.5) -- ++(270:3.5) node [below] {\Large{$Y$}};
\draw[-triangle 90, dotted, very thick] (\x+3.5, \y-1.5) -- ++(270:2.5) node [below] {\Large{$0$}};
\draw[-triangle 90, blue, dotted, very thick] (\x+4.5, \y) -- ++(270:4) node [below] {\Large{$N$}};
\draw[-triangle 90, dotted, very thick] (\x+5.5, \y-1.5) -- ++(270:2.5) node [below] {\Large{$0$}};
\draw[red, dotted, very thick] (\x+6.5, \y) -- ++(270:0.5);
\draw[-triangle 90, blue, dotted, very thick] (\x+6.5, \y-0.5) -- ++(270:3.5) node [below] {\Large{$N$}};
\draw[-triangle 90, dotted, very thick] (\x+7.5, \y-1.5) -- ++(270:2.5) node [below] {\Large{$0$}};
}}

\end{tikzpicture}
}
\end{minipage}
\begin{minipage}{0.05\linewidth}
\ \\
\end{minipage}
\begin{minipage}{0.45\linewidth}
\scalebox{0.6}{
\begin{tikzpicture}
\foreach \x in {0}{
\foreach \y in {8,9,10,11}{
\draw[thick] (\x, \y)-- ++(0:8);
}}
\foreach \x in {0,1,2,3,4,5,6,7,8}{
\foreach \y in {8}{
\draw[thick] (\x, \y)-- ++(90:3);
}}

\foreach \x in {1}{
\foreach \y in {11.25}{
\draw[thick, blue] (\x-0.5,\y) node [above] {\Large{$f_1=N$}};
\draw[thick, orange] (\x+1.5,\y) node [above] {\Large{$f_2=Y^\prime$}};
\draw[thick, blue] (\x+3.5,\y) node [above] {\Large{$f_3=N$}};
\draw[thick, red] (\x+5.5,\y) node [above] {\Large{$f_4=Y$}};
}}

\foreach \x in {0,2,4,6}{
\foreach \y in {8,9,10,11}{
\draw[thick, -open triangle 45] (\x+0.5,\y) ++ (90:0.25) -- ++ (270:0.5);
}}

\foreach \x in {1,3,7,7}{
\foreach \y in {8}{
\draw[thick, -open triangle 45] (\x+0.5,\y) ++ (90:0.25) -- ++ (270:0.5);
}}

\foreach \x in {0}{
\foreach \y in {10}{
\draw[-latex] (\x, \y+0.75)++(0:0.25)-- ++(180:0.5); 
\draw (\x+0.5, \y+0.5) node {\Large{$E$}};
\draw[-latex] (\x+1, \y+0.75)++(0:0.25)-- ++(180:0.5); 
\draw[-latex] (\x+2, \y+0.75)++(0:0.25)-- ++(180:0.5); 
\draw (\x+2.5, \y+0.5) node {\Large{$E$}};
\draw[-latex] (\x+3, \y+0.75)++(0:0.25)-- ++(180:0.5); 
\draw[-latex] (\x+4, \y+0.75)++(0:0.25)-- ++(180:0.5); 
\draw (\x+4.5, \y+0.5) node {\Large{$E$}};
\draw[-latex] (\x+5, \y+0.75)++(0:0.25)-- ++(180:0.5); 
\draw[-latex] (\x+6, \y+0.75)++(0:0.25)-- ++(180:0.5); 
\draw (\x+6.5, \y+0.5) node {\Large{$E$}};
\draw[-latex] (\x+7, \y+0.25)++(0:0.25)-- ++(180:0.5); 
\draw[-latex] (\x+8, \y+0.25)++(0:0.25)-- ++(180:0.5); 
\draw[thick] (\x+8, \y+2)-- ++(0:1)-- ++(270:2)-- ++(180:1)-- ++(90:2);
}}

\foreach \x in {0}{
\foreach \y in {9}{
\draw[thick] (\x-1, \y+2)-- ++(0:1)-- ++(270:2)-- ++(180:1)-- ++(90:2);
\draw[-latex] (\x, \y+0.75)++(180:0.25)-- ++(0:0.5); 
\draw (\x+0.5, \y+0.5) node {\Large{$P_{zig}$}};
\draw[-latex] (\x+1, \y+0.75)++(180:0.25)-- ++(0:0.5); 
\draw (\x+1.5, \y+0.5) node {\Large{$P_{zig}$}};
\draw[-latex] (\x+2, \y+0.75)++(180:0.25)-- ++(0:0.5); 
\draw (\x+2.5, \y+0.5) node {\Large{$P_{zig}$}};
\draw[-latex] (\x+3, \y+0.75)++(180:0.25)-- ++(0:0.5); 
\draw (\x+3.5, \y+0.5) node {\Large{$P_{zig}$}};
\draw[-latex] (\x+4, \y+0.75)++(180:0.25)-- ++(0:0.5); 
\draw (\x+4.5, \y+0.5) node {\Large{$P_{zig}$}};
\draw[-latex] (\x+5, \y+0.75)++(180:0.25)-- ++(0:0.5); 
\draw (\x+5.5, \y+0.5) node {\Large{$P_{zig}$}};
\draw[-latex] (\x+6, \y+0.75)++(180:0.25)-- ++(0:0.5); 
\draw (\x+6.5, \y+0.5) node {\Large{$P_{zig}$}};
\draw[-latex] (\x+7, \y+0.75)++(180:0.25)-- ++(0:0.5); 
\draw (\x+7.5, \y+0.5) node {\Large{$P_{zig}$}};
\draw[-latex] (\x+8, \y+0.75)++(180:0.25)-- ++(0:0.5); 
}}

\foreach \x in {0}{
\foreach \y in {8}{
\draw[thick] (\x-1, \y+1)-- ++(0:1)-- ++(270:2)-- ++(180:1)-- ++(90:2);
\draw[-latex] (\x, \y+0.25)++(0:0.25)-- ++(180:0.5); 
\draw (\x+0.5, \y+0.5) node {\Large{$P_{zag}$}};
\draw[-latex] (\x+1, \y+0.25)++(0:0.25)-- ++(180:0.5); 
\draw (\x+1.5, \y+0.5) node {\Large{$P_{zag}$}};
\draw[-latex] (\x+2, \y+0.25)++(0:0.25)-- ++(180:0.5); 
\draw (\x+2.5, \y+0.5) node {\Large{$P_{zag}$}};
\draw[-latex] (\x+3, \y+0.25)++(0:0.25)-- ++(180:0.5); 
\draw (\x+3.5, \y+0.5) node {\Large{$P_{zag}$}};
\draw[-latex] (\x+4, \y+0.25)++(0:0.25)-- ++(180:0.5); 
\draw (\x+4.5, \y+0.5) node {\Large{$P_{zag}$}};
\draw[-latex] (\x+5, \y+0.25)++(0:0.25)-- ++(180:0.5); 
\draw (\x+5.5, \y+0.5) node {\Large{$P_{zag}$}};
\draw[-latex] (\x+6, \y+0.25)++(0:0.25)-- ++(180:0.5); 
\draw (\x+6.5, \y+0.5) node {\Large{$P_{zag}$}};
\draw[-latex] (\x+7, \y+0.25)++(0:0.25)-- ++(180:0.5); 
\draw (\x+7.5, \y+0.5) node {\Large{$P_{zag}$}};
\draw[-latex] (\x+8, \y+0.25)++(0:0.25)-- ++(180:0.5); 
\draw[thick] (\x+8, \y+2)-- ++(0:1)-- ++(270:2)-- ++(180:1)-- ++(90:2);
}}

\foreach \x in {0}{
\foreach \y in {11}{
\draw[-triangle 90, blue, dotted, very thick] (\x+0.5, \y) -- ++(270:4) node [below] {\Large{$N$}};
\draw[-triangle 90, dotted, very thick] (\x+1.5, \y-1.5) -- ++(270:2.5) node [below] {\Large{$0$}};
\draw[orange, dotted, very thick] (\x+2.5, \y) -- ++(270:0.5);
\draw[-triangle 90, blue, dotted, very thick] (\x+2.5, \y-0.5) -- ++(270:3.5) node [below] {\Large{$N$}};
\draw[-triangle 90, dotted, very thick] (\x+3.5, \y-1.5) -- ++(270:2.5) node [below] {\Large{$0$}};
\draw[-triangle 90, blue, dotted, very thick] (\x+4.5, \y) -- ++(270:4) node [below] {\Large{$N$}};
\draw[-triangle 90, dotted, very thick] (\x+5.5, \y-1.5) -- ++(270:2.5) node [below] {\Large{$0$}};
\draw[red, dotted, very thick] (\x+6.5, \y) -- ++(270:0.5);
\draw[-triangle 90, red, dotted, very thick] (\x+6.5, \y-0.5) -- ++(270:3.5) node [below] {\Large{$Y$}};
\draw[-triangle 90, dotted, very thick] (\x+7.5, \y-1.5) -- ++(270:2.5) node [below] {\Large{$0$}};
}}

\end{tikzpicture}
}
\end{minipage}

\caption{Example run of Module 3, in which the transitions that have proved valid in Modules 1 and 2 ($f_2$ and $f_4$ here) are chosen nondeterministically.}
\label{fig:example_module3}
\end{figure}

Let us assume that the transcript has been folded up to its $(i{-}1)$-th period successfully into a parallelogram, which outputs the state $q_{i-1}$ below, and the next period reads the letter $b_i$.  
See Figures~\ref{fig:example_FA_module1}, \ref{fig:example_module2}, \ref{fig:example_module3}, and \ref{fig:example_module4} for an example run. 

\paragraph{Bricks.}
All the modules (more precisely, their transcripts) consist of functional submodules. 
Each submodule expects several surrounding environments. 
A conformation that the submodule takes in such an expected environment is called a \textit{brick} \cite{GeMeScSe2018}. 
On the inductive assumption that all the previous submodules have folded into a brick, a submodule never encounters an unexpected environment, and hence, folds into a brick. 
Any expected environment exposes 1-bit information $b$ below and a submodule enters it in such a manner that its first bead is placed either 1-bead below $y$ (\textit{starting at the top}) or 3-beads below $y$ (\textit{at the bottom}). 
Hence, in this paper, a brick of a module $X$ is denoted as $X_{\rm -hy}$, where ${\rm h} \in \{{\rm t}, {\rm b}\}$ indicates whether this brick starts at the top or bottom and y is the input from above. 

\paragraph{Spacers and Turners.}
The transcript of a zig or a zag is a chain of submodules interleaved by a structural sequence called a \textit{spacer}. 
A spacer keeps two continuous submodules far enough horizontally so as to prevent their undesirable interaction. 
We employ spacers that fold into a parallelogram (p-spacer) or glider (g-spacer) of height~3.
For a g-spacer, see Figure~\ref{fig:glider}. 
They start and end folding at the same height (top or bottom) in order to propagate 1-bit of information. 
The spacer and its 1-bit carrying capability are classical, found already in the first oritatami system \cite{GeMeScSe2016}. 

After a zig is transcribed a structural submodule called \textit{turner}. 
Its role is to fold so as to guide the transcript to the point where the next zag is supposed to start. 
A zag is also followed by a turner for the next zig. 
Some turners play also a functional role. 

\begin{figure}[tb]
\centering
\begin{minipage}{0.45\linewidth}
\centering
\includegraphics[width=\linewidth]{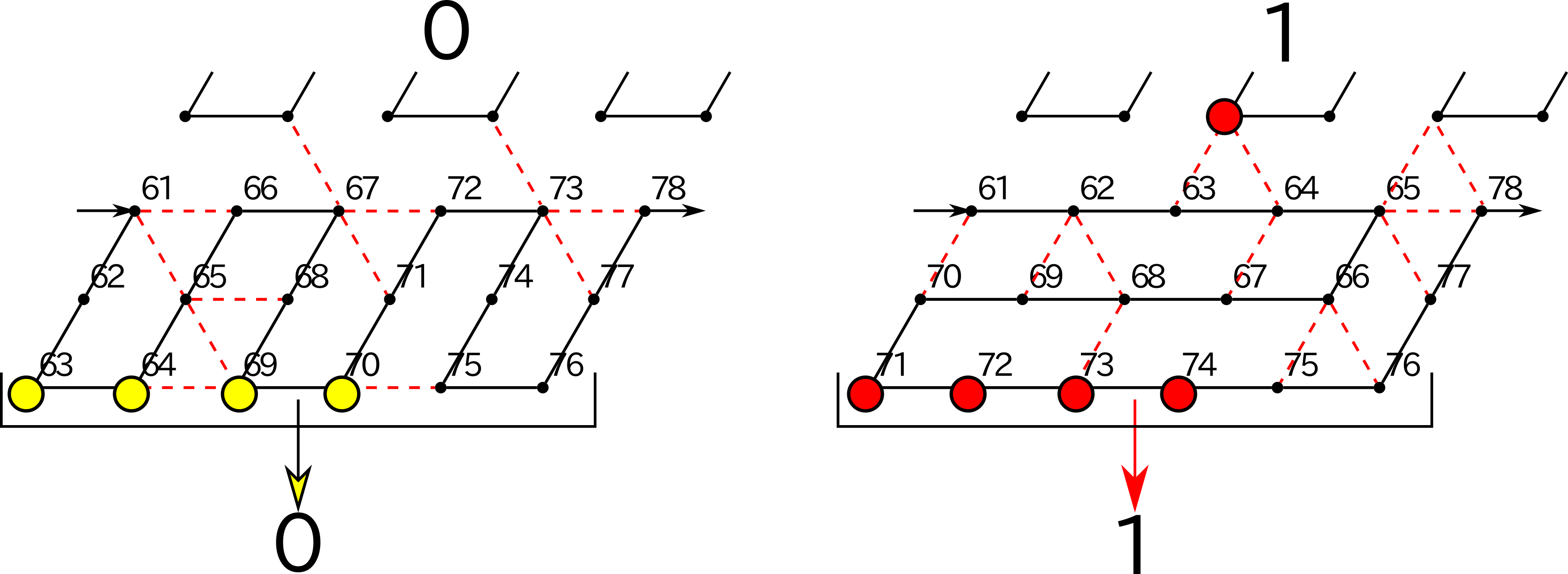}
\end{minipage}
\begin{minipage}{0.05\linewidth}
\ \\
\end{minipage}
\begin{minipage}{0.45\linewidth}
\centering
\includegraphics[width=\linewidth]{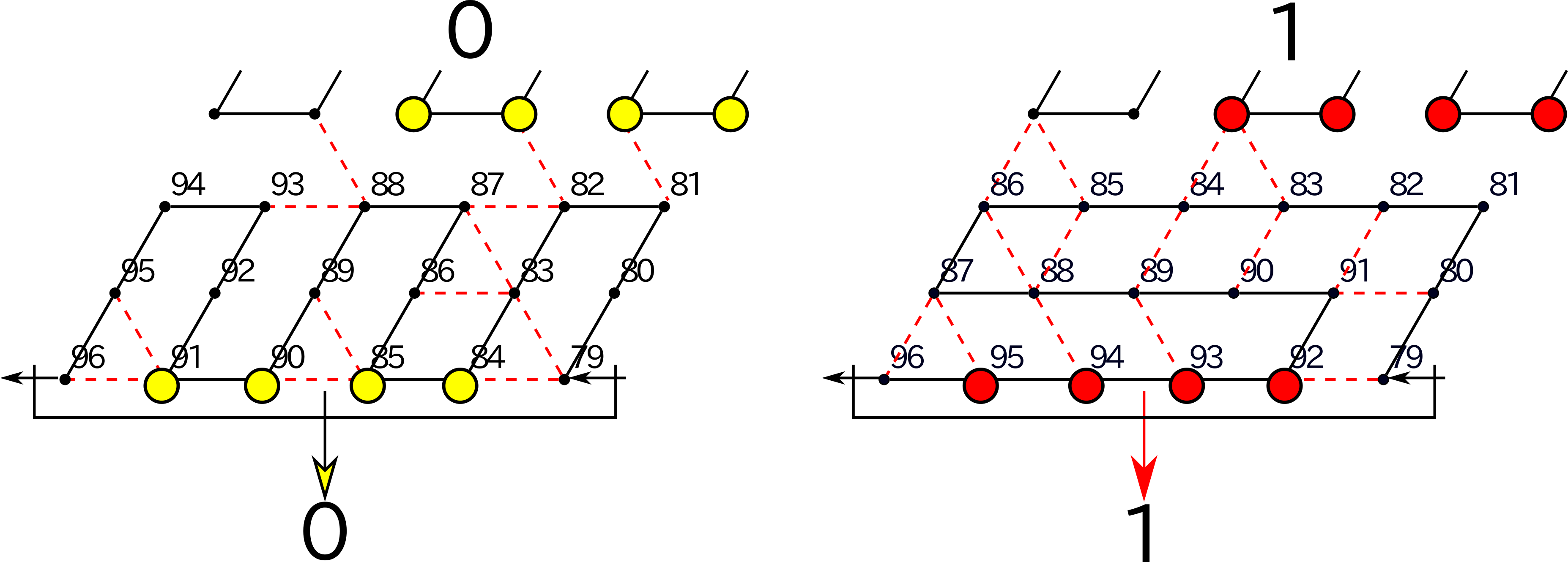}
\end{minipage}
\caption{
(Left) The two bricks of $P_{\rm zig}$, that is, $P_{\rm zig-0t}$ and $P_{\rm zig-1t}$. 
(Right) The two bricks of $P_{\rm zag}$, that is, $P_{\rm zag-0b}$ and $P_{\rm zag-1b}$.}
\label{fig:brick_P}
\end{figure}

\paragraph{Module 1 (origin state checker)} folds into $2n$ zigzags. 
Recall that all the $n$ variables $f_1, f_2, \ldots, f_n$ have been set to $N$ by the seed or Module~4 in the previous period. 
The $(2k{-}1)$-th zigzag checks whether the origin $o_k$ of the $k$-th transition $f_k$ is equal to $q_{i-1}$ or not, and if so, it sets the variable $f_k$ to $Y$. 
Every other zigzag (2nd, 4th, and so on) just formats these variables as well as the $z$-variables (for the current state in \eqref{format:state}) using two submodules $P_{\rm zig}$ and $P_{\rm zag}$ (see Figure~\ref{fig:brick_P} for their bricks); this is a common feature among all the four modules. 
The transcript for such a \textit{formatting} zig (resp. zag) is a chain of $2n$ instances of $P_{\rm zig}$ (resp.~$P_{\rm zag}$), unless otherwise noted. 

\begin{figure}[tb]
\centering
\includegraphics[width=\linewidth]{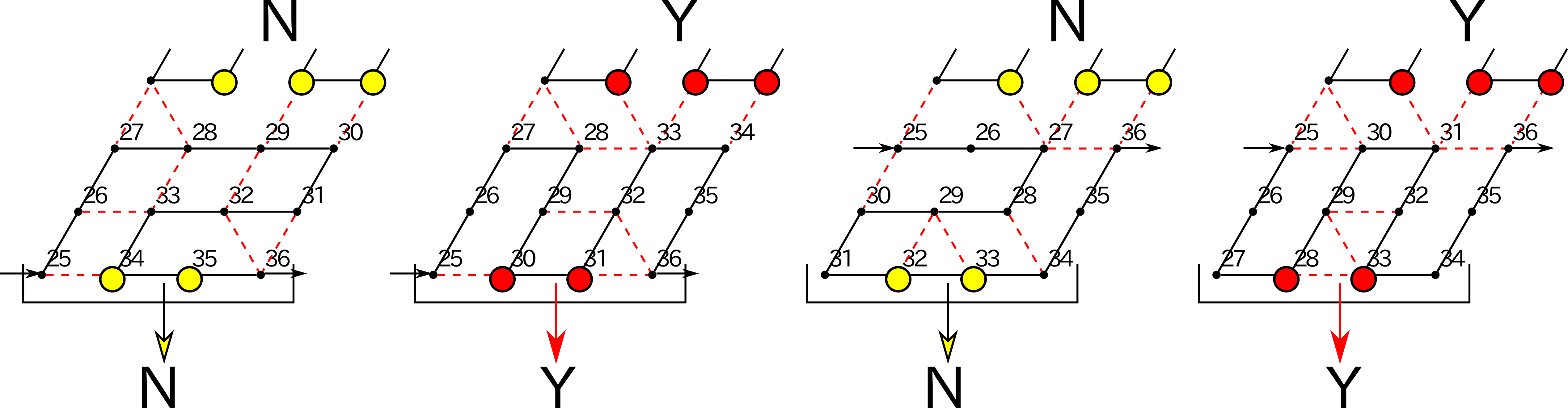}
\caption{The four bricks of $A'$, that is, $A'_{\rm Nb}, A'_{\rm Yb}, A'_{\rm Nt}$, and $A'_{\rm Yt}$.}
\label{fig:brick_A'}
\end{figure}

\begin{figure}[tb]
\centering
\includegraphics[width=\linewidth]{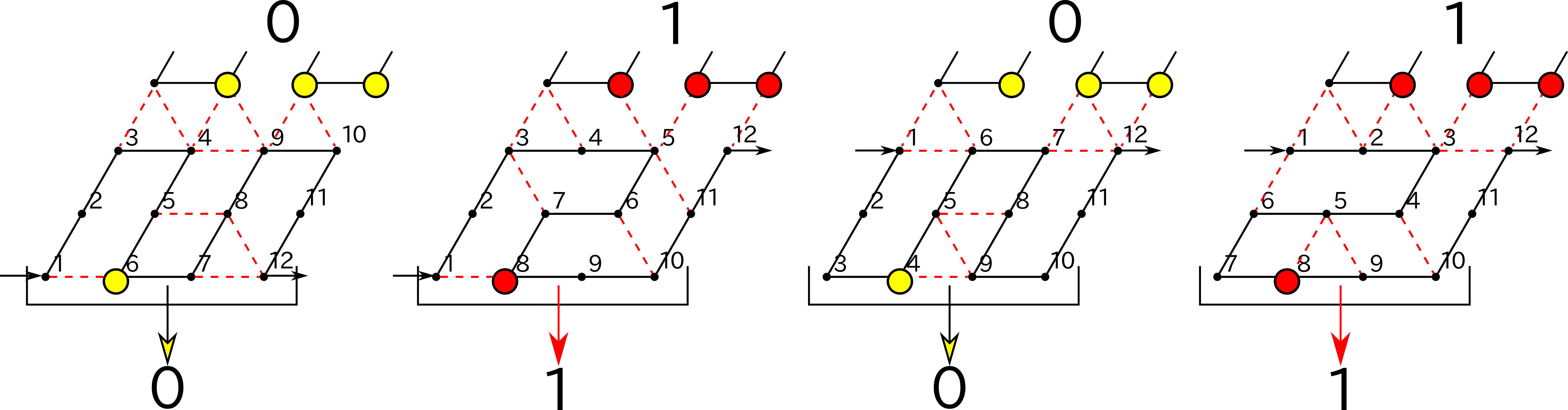}
\caption{The four bricks of $A_0$, that is, $A_{\rm 0-0b}, A_{\rm 0-1b}, A_{\rm 0-0t}$, and $A_{\rm 0-1t}$}
\label{fig:brick_A0}
\end{figure}

\begin{figure}[tb]
\centering
\includegraphics[width=\linewidth]{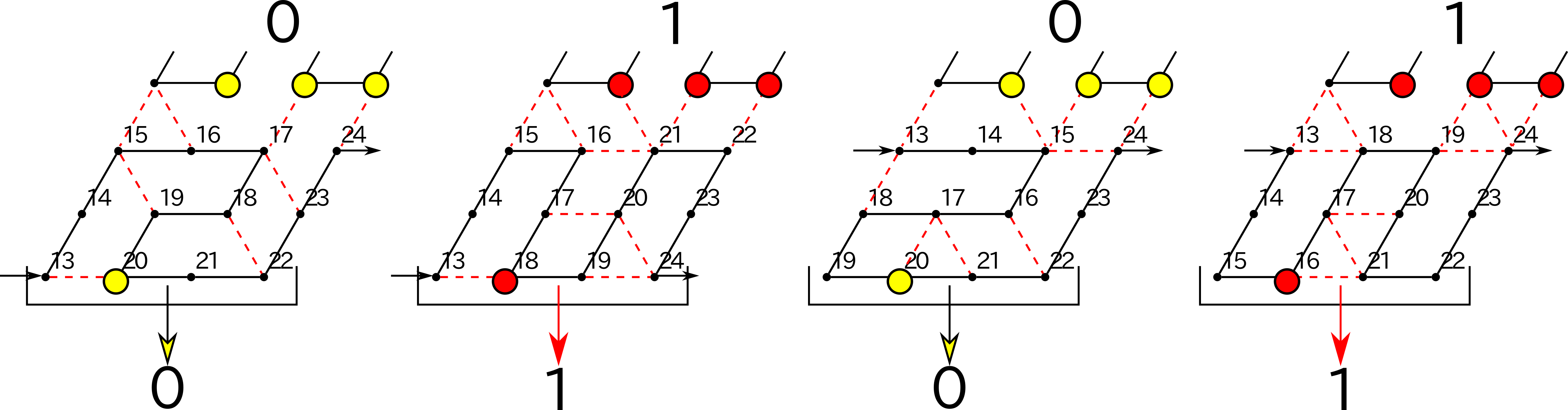}
\caption{The four bricks of $A_1$, that is, $A_{\rm 1-0b}, A_{\rm 1-1b}, A_{\rm 1-0t}$, and $A_{\rm 1-1t}$.}
\label{fig:brick_A1}
\end{figure}

The transcript for the $(2k{-}1)$-th zig is semantically represented as $\odot_{j = 1}^n (A' A_{o_k[j]})$ for submodules $A', A_0, A_1$. 
See Figures~\ref{fig:brick_A'}, \ref{fig:brick_A0}, and \ref{fig:brick_A1} for the four bricks of these submodules, respectively. 
The zig starts folding at the bottom. 
The $n$ instances of $A'$ propagate $f_1, \ldots, f_n$ downward using the four bricks, all of which end folding at the same height as they start.
$A_{o_k[j]}$ checks whether $o_k[j] = q_{i-1}[j]$ or not when it starts at the bottom; it ends at the bottom if these bits are equal, or top otherwise. 
Starting at the top, it certainly ends at the top. 
In any case, it propagates $q_{i-1}[j]$ downward.  
The zig thus ends at the bottom iff $o_k = q_{i-1}$. 
The succeeding turner admits two conformations to let the next zag start either at the bottom if $o_k = q_{i-1}$, or top otherwise. 

\begin{figure}[tb]
\centering
\includegraphics[width=\linewidth]{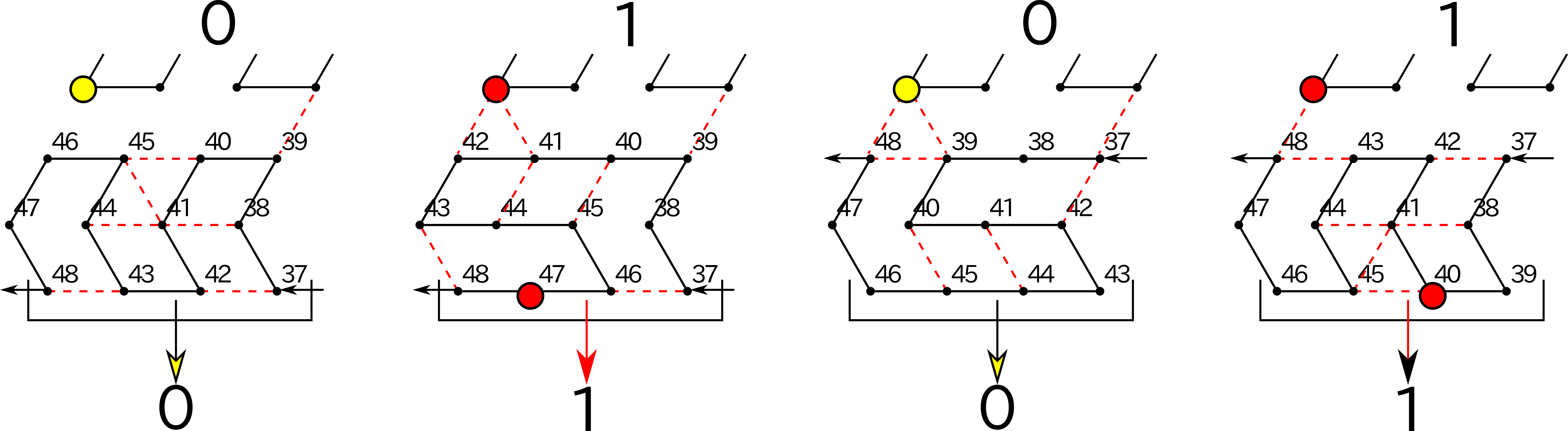}
\caption{The four bricks of $B$, that is, $B_{\rm 0b}, B_{\rm 1b}, B_{\rm 0t}$, and $B_{\rm 1t}$.}
\label{fig:brick_B}
\end{figure}

The transcript for the next zag is $B^{2n-2k+1} B' B^{2k-2}$ for submodules $B$ (see Figure~\ref{fig:brick_B} for its bricks) and $B'$. 
It is transcribed from right to left so that $B'$ can read $f_k$. 
$B'$ is in fact just a g-spacer shown in Figure~\ref{fig:glider}. 
This glider exposes below the bead-type sequence ${\tt 590}$-${\tt 585}$-${\tt 584}$-${\tt 579}$ if it starts folding at the bottom, or ${\tt 588}$-${\tt 587}$-${\tt 582}$-${\tt 581}$ otherwise; the former and latter shall be formatted into $x_Y$ and $x_N$, respectively, by the next zigzag. 

The variables $f_1, \ldots, f_n$ are updated in this way and output below in the format \eqref{format:state} along with the current state $q_{i-1}$, which is not used any more though. 

\begin{figure}[tb]
\centering
\includegraphics[width=\linewidth]{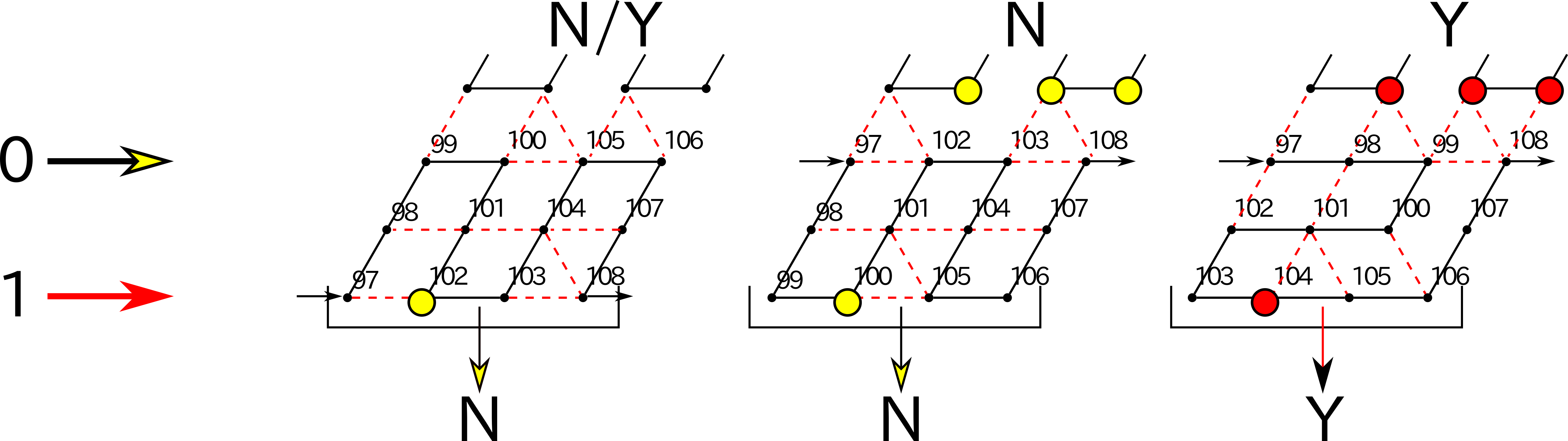}
\caption{The three bricks of $C_0$, that is, $C_{\rm 0-\ast b}, C_{\rm 0-Nt}$, and $C_{\rm 0-Yt}$.} 
\label{fig:brick_C0}
\end{figure}

\begin{figure}[tb]
\centering
\includegraphics[width=\linewidth]{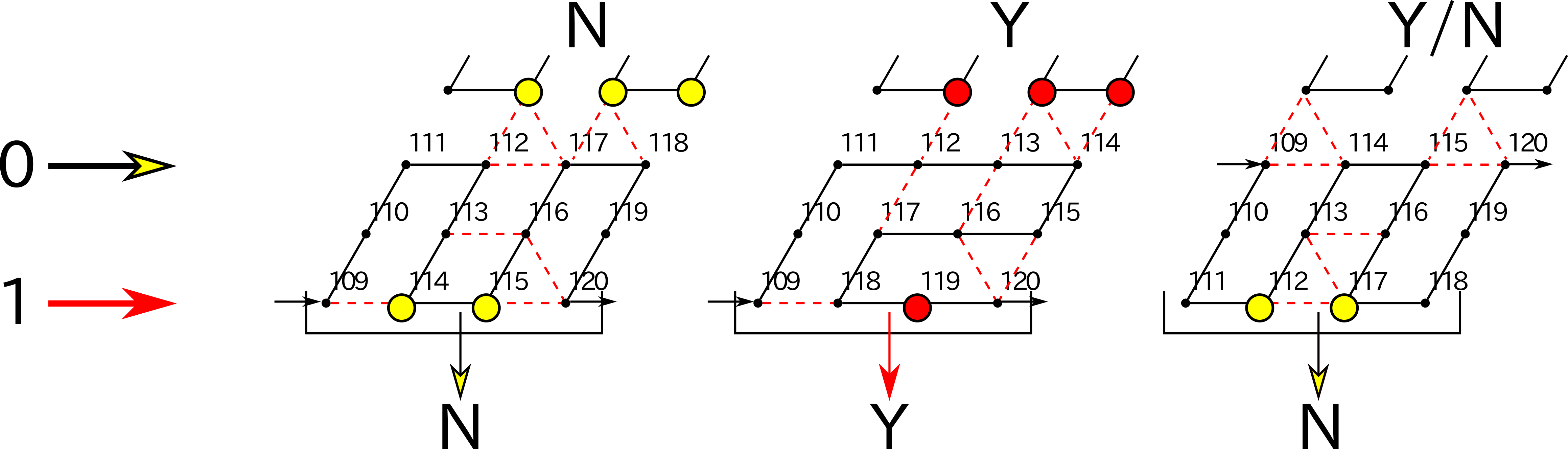}
\caption{The three bricks of $C_1$, that is, $C_{\rm 1-Nb}, C_{\rm 1-Yb}$, and $C_{\rm 1-\ast t}$.}
\label{fig:brick_C1}
\end{figure}

\paragraph{Module 2 (input letter checker)} folds into $2m$ zigzags; recall $m = \lceil \log |\Sigma| \rceil$. 
The $\ell$-th bit of the input letter $b_i$ is read by the turner between the $(2\ell{-}2)$-th zag and $(2\ell{-}1)$-th zig and the bit lets this zig start at the top if it is 0, or bottom if it is 1. 
Recall that $f_k$ reads $a_k$ for all $1 \le k \le n$. 
The $\ell$-th bit of these letters is encoded in the transcript for the $(2\ell{-}1)$-th zig as $C_{a_1[\ell]} C_{a_2[\ell]} \cdots C_{a_n[\ell]}$ using submodules $C_0$ and $C_1$. 
All the bricks of $C_0$ and $C_1$ start and end at the same height, as shown in Figures~\ref{fig:brick_C0} and \ref{fig:brick_C1}; thus propagating $b_i[\ell]$ throughout the zig.  
Starting at the top (i.e., $b_i[\ell] = 0$), $C_0$ takes the brick $C_{\rm 0-Nt}$ if it reads $N$ from above or $C_{\rm 0-Yt}$ if it reads $Y$; these bricks output $N$ and $Y$ downward, respectively; thus propagating the $x$-variables downward. 
On the other hand, if it starts at the bottom (i.e., $b_i[\ell] = 1$), $C_0$ certainly takes $C_{\rm 0-\ast b}$ and outputs $N$ downward. 
$C_1$ propagates what it reads downward by the bricks $C_{\rm 1-Nb}, C_{\rm 1-Yb}$ in Figure~\ref{fig:brick_C1} if $b_i[\ell] = 1$ while it outputs $N$ downward by $C_{\rm 1-\ast t}$ if $b_i[\ell] = 0$. 
Functioning in this way, the submodules $C_{a_1[j]}, \ldots, C_{a_n[j]}$ compare the letters that $f_1, \ldots, f_n$ read with $b_i$ and filter those with unmatching $j$-th bit out. 
The next zag propagates the result of this filtering downward using $B$'s. 

\begin{figure}[tb]
\centering
\includegraphics[width=0.8\linewidth]{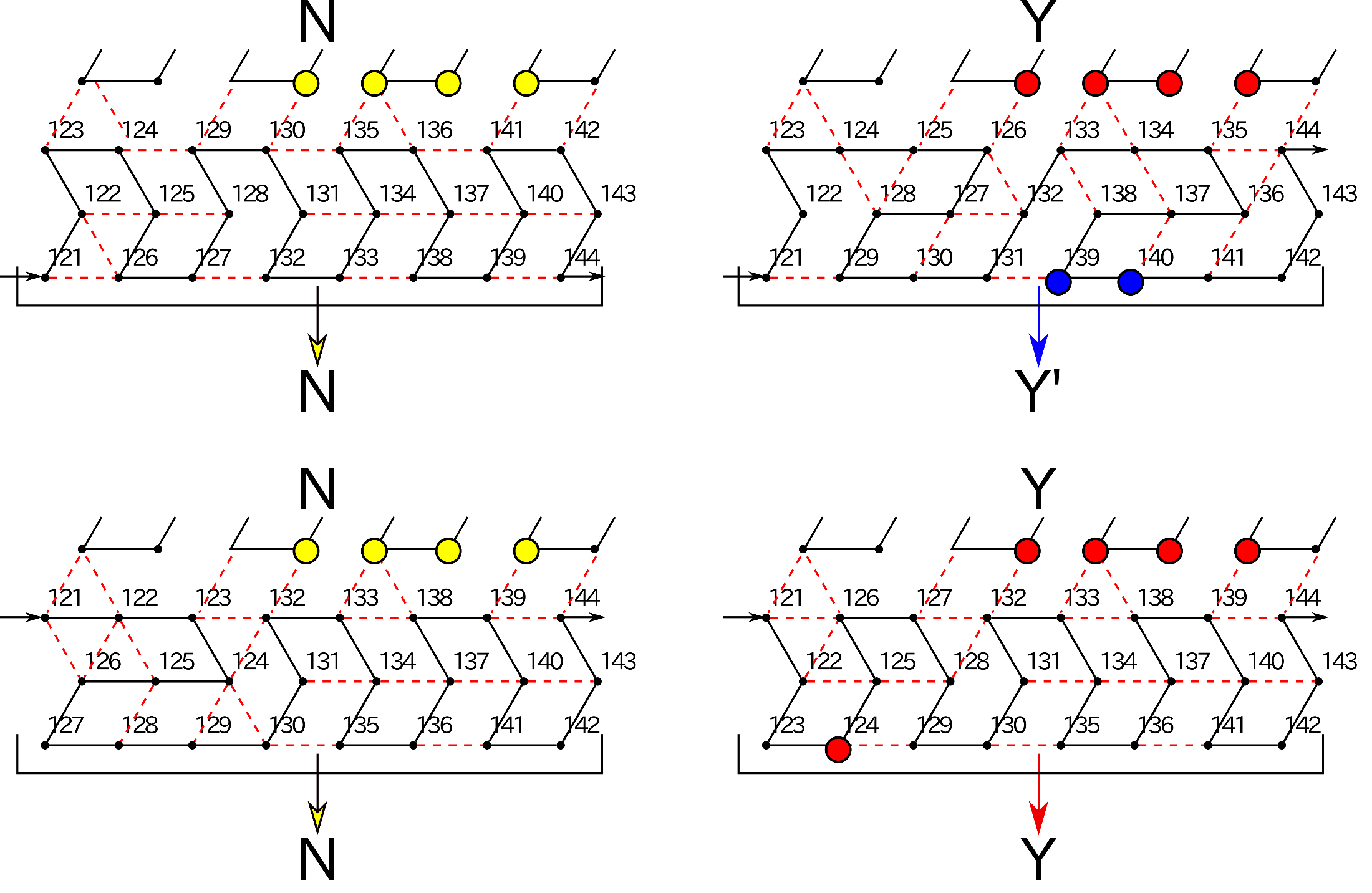}
\caption{The four bricks of $D$: (Top) $D_{\rm -Nb}$ and $D_{\rm -Yb}$; (Bottom) $D_{\rm -Nt}$ and $D_{\rm -Yt}$.}
\label{fig:brick_D}
\end{figure}

\begin{figure}[tb]
\centering
\includegraphics[width=\linewidth]{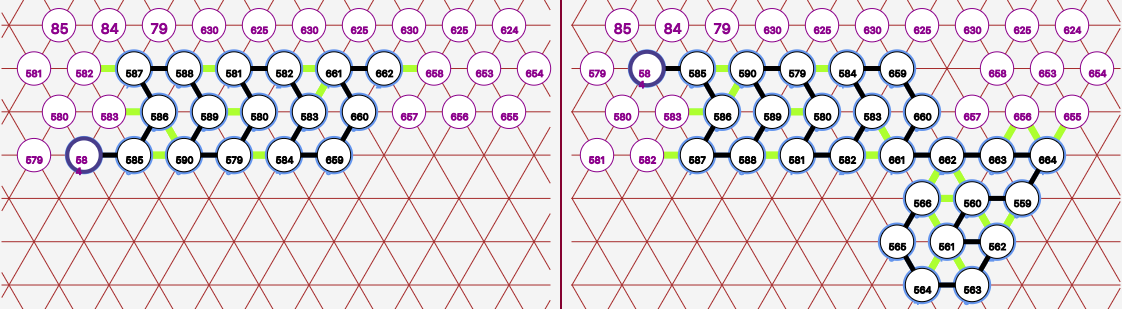}
\caption{Turner from the first zig of Module 3 to the first zag. 
(Left) It is trapped geometrically in the pocket of the previous turner and halts the system if it starts folding at the bottom. 
(Right) It is not trapped and lets the next zag be transcribed.}
\label{fig:module3_halt}
\end{figure}

\paragraph{Module 3 (nondeterministic choice of the next transition)} folds into just 2 zigzags. 
Each transition $f_k = (o_k, a_k, t_k)$ has been checked whether $o_k = q_{i-1}$ in Module 1 and whether $a_k = b_i$ in Module 2, and the variable $f_k$ is set to $Y$ iff $f_k$ passed both the checks, that is, proved \textit{valid}. 
The first zig marks the valid transition with smallest subscript by setting its variable to $Y'$ using a submodule $D$. 
This submodule was invented in \cite{MasudaSekiUbukata2018} for the same purpose, and hence, we just mention a property that its four bricks (Figure~\ref{fig:brick_D}) ensure this zig to end at the bottom if none of the transition has proven valid. 
In that case, the succeeding turner is geometrically trapped as shown in Figure~\ref{fig:module3_halt} and the system halts. 

\begin{figure}[tb]
\centering
\includegraphics[width=\linewidth]{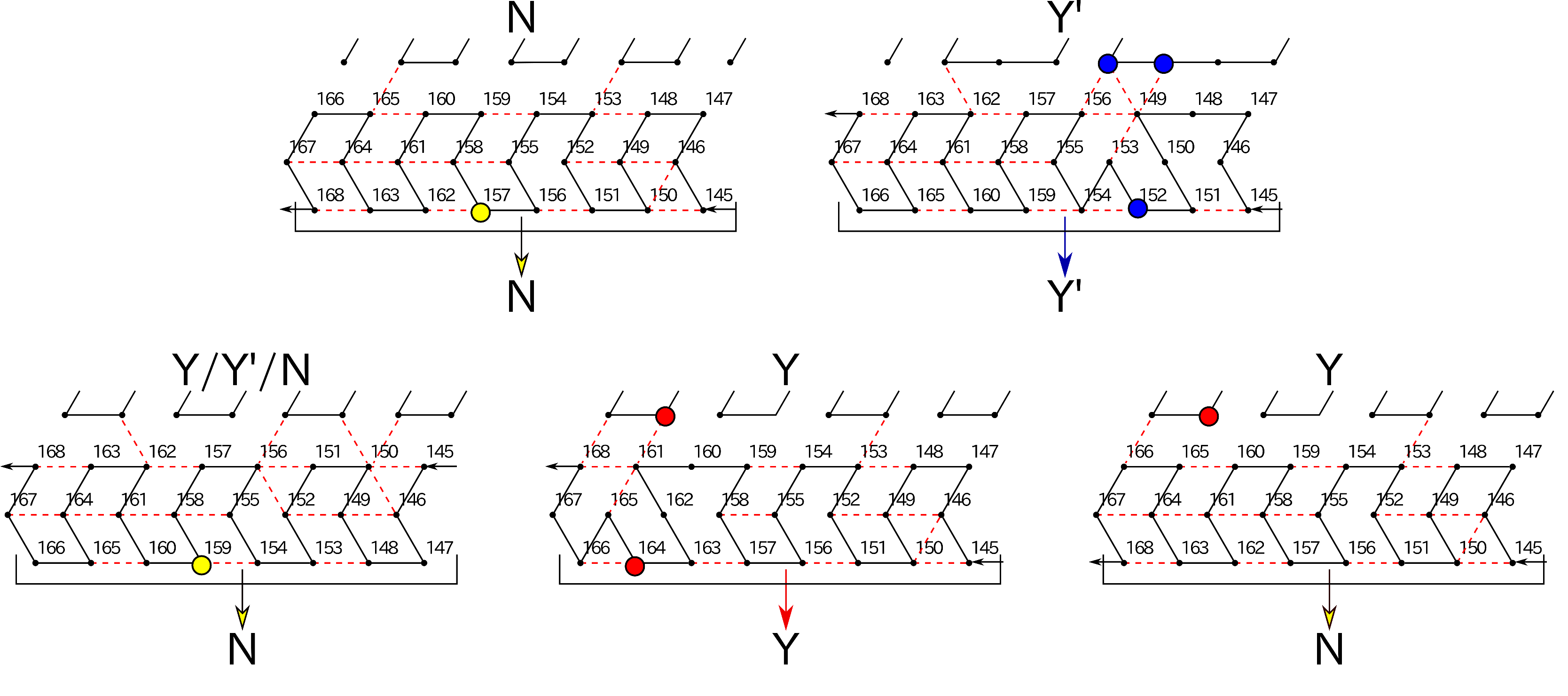}
\caption{The five bricks of $E$: (Top) $E_{\rm -Nb}$ and $E_{\rm -Y'b}$; (Bottom) $E_{\rm -\ast t}, E_{\rm -YbY}$, and $E_{\rm -YbN}$. 
Note that $E_{\rm -YbY}$ and $E_{\rm -YbN}$ are chosen nondeterministically and equally probably.}
\label{fig:brick_E}
\end{figure}

The transcript for the first zag consists of $n$ instances of a submodule $E$. 
The five bricks of $E$ are shown in Figure~\ref{fig:brick_E}. 
The zag starts folding at the bottom. 
When an $E$ starts at the bottom and reads $Y$ from above, it folds into the brick $E_{\rm -YbY}$ or $E_{\rm -YbN}$ in Figure~\ref{fig:brick_E} nondeterministically, which amounts to choosing the corresponding valid transition or not. 
Observe that $E_{\rm -YbY}$ ends at the top, notifying the succeeding $E$'s that the decision has been already made. 
When starting at the top, $E$ takes no brick but $E_{\rm -\ast t}$, which outputs $N$ no matter what it reads. 
The brick $E_{\rm -Y'b}$ and $Y'$ (marked $Y$) prevent the oritatami system from not choosing any valid transition; that is, if an $E$ starts at the bottom (meaning that none of the valid transitions has been chosen yet) and reads $Y'$ from above, it deterministically folds into $E_{\rm -Y'b}$, which outputs $Y$. 

The transcript of the next zig differs from that of normal formatting zig in that every other instance of $P_{\rm zig}$ is replaced by a spacer. 
This replacement allows the $n$ instances of $P_{\rm zag}$ responsible for the $z$-variables to take their ``default'' brick $P_{\rm zag-0b}$, which outputs 0. 
This is a preprocess for Module 4 to set these variables to the target state of the transition chosen. 

\begin{figure}[tb]
\centering
\scalebox{0.8}{\begin{tikzpicture}
\foreach \x in {0}{
\foreach \y in {5,6,7,8,9,10,11,12,13,14,15}{
\draw[thick] (\x, \y)-- ++(0:8);
}}
\foreach \x in {0,1,2,3,4,5,6,7,8}{
\foreach \y in {5}{
\draw[thick] (\x, \y)-- ++(90:10);
}}

\foreach \x in {1}{
\foreach \y in {15.25}{
\draw[thick, blue] (\x-0.5,\y) node [above] {\LARGE{$N$}};
\draw[thick] (\x+0.5,\y) node [above] {\LARGE{$0$}};
\draw[thick, red] (\x+1.5,\y) node [above] {\LARGE{$Y$}};
\draw[thick] (\x+2.5,\y) node [above] {\LARGE{$0$}};
\draw[thick, blue] (\x+3.5,\y) node [above] {\LARGE{$N$}};
\draw[thick] (\x+4.5,\y) node [above] {\LARGE{$0$}};
\draw[thick, blue] (\x+5.5,\y) node [above] {\LARGE{$N$}};
\draw[thick] (\x+6.5,\y) node [above] {\LARGE{$0$}};
}}

\foreach \x in {0,1,2,3,4,5,6,7}{
\foreach \y in {5,6,7,8,9,10,11,12,13,14,15}{
\draw[thick, -open triangle 45] (\x+0.5,\y) ++ (90:0.25) -- ++ (270:0.5);
}}

\foreach \x in {0}{
\foreach \y in {14}{
\draw[-latex] (\x, \y+0.25)++(180:0.25)-- ++(0:0.5);
\draw (\x+0.5, \y+0.5) node {\Large{$A_1$}};
\draw[-latex] (\x+1, \y+0.75)++(180:0.25)-- ++(0:0.5);
\draw (\x+1.5, \y+0.5) node {\Large{$A^\prime$}};
\draw[-latex] (\x+2, \y+0.75)++(180:0.25)-- ++(0:0.5);
\draw (\x+2.5, \y+0.5) node {\Large{$A^\prime$}};
\draw[-latex] (\x+3, \y+0.75)++(180:0.25)-- ++(0:0.5);
\draw (\x+3.5, \y+0.5) node {\Large{$A^\prime$}};
\draw[-latex] (\x+4, \y+0.75)++(180:0.25)-- ++(0:0.5);
\draw (\x+4.5, \y+0.5) node {\Large{$A^\prime$}};
\draw[-latex] (\x+5, \y+0.75)++(180:0.25)-- ++(0:0.5);
\draw (\x+5.5, \y+0.5) node {\Large{$A^\prime$}};
\draw[-latex] (\x+6, \y+0.75)++(180:0.25)-- ++(0:0.5);
\draw (\x+6.5, \y+0.5) node {\Large{$A^\prime$}};
\draw[-latex] (\x+7, \y+0.75)++(180:0.25)-- ++(0:0.5);
\draw (\x+7.5, \y+0.5) node {\Large{$A^\prime$}};
\draw[-latex] (\x+8, \y+0.75)++(180:0.25)-- ++(0:0.5);
\draw[thick] (\x+8, \y+1)-- ++(0:1)-- ++(270:2)-- ++(180:1);
}}

\foreach \x in {0}{
\foreach \y in {13}{
\draw[thick] (\x, \y+1)-- ++(180:1)-- ++(270:2)-- ++(0:1);
\draw[-latex] (\x, \y+0.75)++(0:0.25)-- ++(180:0.5); 
\draw (\x+0.5, \y+0.5) node {\Large{$G_0$}};
\draw[-latex] (\x+1, \y+0.75)++(0:0.25)-- ++(180:0.5); 
\draw (\x+1.5, \y+0.5) node {\Large{$G_0$}};
\draw[-latex] (\x+2, \y+0.75)++(0:0.25)-- ++(180:0.5); 
\draw (\x+2.5, \y+0.5) node {\Large{$B$}};
\draw[-latex] (\x+3, \y+0.75)++(0:0.25)-- ++(180:0.5); 
\draw (\x+3.5, \y+0.5) node {\Large{$G_0$}};
\draw[-latex] (\x+4, \y+0.75)++(0:0.25)-- ++(180:0.5); 
\draw (\x+4.5, \y+0.5) node {\Large{$B$}};
\draw[-latex] (\x+5, \y+0.75)++(0:0.25)-- ++(180:0.5); 
\draw (\x+5.5, \y+0.5) node {\Large{$G_1$}};
\draw[-latex] (\x+6, \y+0.75)++(0:0.25)-- ++(180:0.5); 
\draw (\x+6.5, \y+0.5) node {\Large{$B$}};
\draw[-latex] (\x+7, \y+0.75)++(0:0.25)-- ++(180:0.5); 
\draw (\x+7.5, \y+0.5) node {\Large{$G_1$}};
\draw[-latex] (\x+8, \y+0.75)++(0:0.25)-- ++(180:0.5); 
}}

\foreach \x in {0}{
\foreach \y in {10}{
\draw[-latex] (\x, \y+0.25)++(180:0.25)-- ++(0:0.5);
\draw (\x+0.5, \y+0.5) node {\Large{$A^\prime$}};
\draw[-latex] (\x+1, \y+0.25)++(180:0.25)-- ++(0:0.5);
\draw (\x+1.5, \y+0.5) node {\Large{$A^\prime$}};
\draw[-latex] (\x+2, \y+0.25)++(180:0.25)-- ++(0:0.5);
\draw (\x+2.5, \y+0.5) node {\Large{$A_1$}};
\draw[-latex] (\x+3, \y+0.25)++(180:0.25)-- ++(0:0.5);
\draw (\x+3.5, \y+0.5) node {\Large{$A^\prime$}};
\draw[-latex] (\x+4, \y+0.25)++(180:0.25)-- ++(0:0.5);
\draw (\x+4.5, \y+0.5) node {\Large{$A^\prime$}};
\draw[-latex] (\x+5, \y+0.25)++(180:0.25)-- ++(0:0.5);
\draw (\x+5.5, \y+0.5) node {\Large{$A^\prime$}};
\draw[-latex] (\x+6, \y+0.25)++(180:0.25)-- ++(0:0.5);
\draw (\x+6.5, \y+0.5) node {\Large{$A^\prime$}};
\draw[-latex] (\x+7, \y+0.25)++(180:0.25)-- ++(0:0.5);
\draw (\x+7.5, \y+0.5) node {\Large{$A^\prime$}};
\draw[-latex] (\x+8, \y+0.25)++(180:0.25)-- ++(0:0.5);
\draw[thick] (\x+8, \y+1)-- ++(0:1)-- ++(270:2)-- ++(180:1);
}}

\foreach \x in {0}{
\foreach \y in {9}{
\draw[thick] (\x, \y+1)-- ++(180:1)-- ++(270:2)-- ++(0:1);
\draw[-latex] (\x, \y+0.25)++(0:0.25)-- ++(180:0.5); 
\draw (\x+0.5, \y+0.5) node {\Large{$B$}};
\draw[-latex] (\x+1, \y+0.25)++(0:0.25)-- ++(180:0.5); 
\draw (\x+1.5, \y+0.5) node {\Large{$G_1$}};
\draw[-latex] (\x+2, \y+0.25)++(0:0.25)-- ++(180:0.5); 
\draw (\x+2.5, \y+0.5) node {\Large{$G_0$}};
\draw[-latex] (\x+3, \y+0.25)++(0:0.25)-- ++(180:0.5); 
\draw (\x+3.5, \y+0.5) node {\Large{$G_0$}};
\draw[-latex] (\x+4, \y+0.25)++(0:0.25)-- ++(180:0.5); 
\draw (\x+4.5, \y+0.5) node {\Large{$B$}};
\draw[-latex] (\x+5, \y+0.25)++(0:0.25)-- ++(180:0.5); 
\draw (\x+5.5, \y+0.5) node {\Large{$G_0$}};
\draw[-latex] (\x+6, \y+0.25)++(0:0.25)-- ++(180:0.5); 
\draw (\x+6.5, \y+0.5) node {\Large{$B$}};
\draw[-latex] (\x+7, \y+0.25)++(0:0.25)-- ++(180:0.5); 
\draw (\x+7.5, \y+0.5) node {\Large{$G_0$}};
\draw[-latex] (\x+8, \y+0.25)++(0:0.25)-- ++(180:0.5); 
}}

\foreach \x in {0}{
\foreach \y in {6}{
\draw[-latex] (\x, \y+0.25)++(180:0.25)-- ++(0:0.5);
\draw (\x+0.5, \y+0.5) node {\Large{$A^\prime$}};
\draw[-latex] (\x+1, \y+0.25)++(180:0.25)-- ++(0:0.5);
\draw (\x+1.5, \y+0.5) node {\Large{$A^\prime$}};
\draw[-latex] (\x+2, \y+0.25)++(180:0.25)-- ++(0:0.5);
\draw (\x+2.5, \y+0.5) node {\Large{$A^\prime$}};
\draw[-latex] (\x+3, \y+0.25)++(180:0.25)-- ++(0:0.5);
\draw (\x+3.5, \y+0.5) node {\Large{$A^\prime$}};
\draw[-latex] (\x+4, \y+0.25)++(180:0.25)-- ++(0:0.5);
\draw (\x+4.5, \y+0.5) node {\Large{$A_1$}};
\draw[-latex] (\x+5, \y+0.75)++(180:0.25)-- ++(0:0.5);
\draw (\x+5.5, \y+0.5) node {\Large{$A^\prime$}};
\draw[-latex] (\x+6, \y+0.75)++(180:0.25)-- ++(0:0.5);
\draw (\x+6.5, \y+0.5) node {\Large{$A^\prime$}};
\draw[-latex] (\x+7, \y+0.75)++(180:0.25)-- ++(0:0.5);
\draw (\x+7.5, \y+0.5) node {\Large{$A^\prime$}};
\draw[-latex] (\x+8, \y+0.75)++(180:0.25)-- ++(0:0.5);
\draw[thick] (\x+8, \y+1)-- ++(0:1)-- ++(270:2)-- ++(180:1);
}}

\foreach \x in {0}{
\foreach \y in {5}{
\draw[thick] (\x, \y+1)-- ++(180:1)-- ++(270:2)-- ++(0:1);
\draw[-latex] (\x, \y+0.75)++(0:0.25)-- ++(180:0.5); 
\draw (\x+0.5, \y+0.5) node {\Large{$B$}};
\draw[-latex] (\x+1, \y+0.75)++(0:0.25)-- ++(180:0.5); 
\draw (\x+1.5, \y+0.5) node {\Large{$G_0$}};
\draw[-latex] (\x+2, \y+0.75)++(0:0.25)-- ++(180:0.5); 
\draw (\x+2.5, \y+0.5) node {\Large{$B$}};
\draw[-latex] (\x+3, \y+0.75)++(0:0.25)-- ++(180:0.5); 
\draw (\x+3.5, \y+0.5) node {\Large{$G_0$}};
\draw[-latex] (\x+4, \y+0.75)++(0:0.25)-- ++(180:0.5); 
\draw (\x+4.5, \y+0.5) node {\Large{$G_0$}};
\draw[-latex] (\x+5, \y+0.75)++(0:0.25)-- ++(180:0.5); 
\draw (\x+5.5, \y+0.5) node {\Large{$G_1$}};
\draw[-latex] (\x+6, \y+0.75)++(0:0.25)-- ++(180:0.5); 
\draw (\x+6.5, \y+0.5) node {\Large{$B$}};
\draw[-latex] (\x+7, \y+0.75)++(0:0.25)-- ++(180:0.5); 
\draw (\x+7.5, \y+0.5) node {\Large{$G_1$}};
\draw[-latex] (\x+8, \y+0.75)++(0:0.25)-- ++(180:0.5); 
}}

\foreach \x in {}{
\foreach \y in {}{
\draw[-latex] (\x, \y+0.25)++(180:0.25)-- ++(0:0.5);
\draw (\x+0.5, \y+0.5) node {\Large{$A^\prime$}};
\draw[-latex] (\x+1, \y+0.25)++(180:0.25)-- ++(0:0.5);
\draw (\x+1.5, \y+0.5) node {\Large{$A^\prime$}};
\draw[-latex] (\x+2, \y+0.25)++(180:0.25)-- ++(0:0.5);
\draw (\x+2.5, \y+0.5) node {\Large{$A^\prime$}};
\draw[-latex] (\x+3, \y+0.25)++(180:0.25)-- ++(0:0.5);
\draw (\x+3.5, \y+0.5) node {\Large{$A^\prime$}};
\draw[-latex] (\x+4, \y+0.25)++(180:0.25)-- ++(0:0.5);
\draw (\x+4.5, \y+0.5) node {\Large{$A^\prime$}};
\draw[-latex] (\x+5, \y+0.25)++(180:0.25)-- ++(0:0.5);
\draw (\x+5.5, \y+0.5) node {\Large{$A^\prime$}};
\draw[-latex] (\x+6, \y+0.25)++(180:0.25)-- ++(0:0.5);
\draw (\x+6.5, \y+0.5) node {\Large{$A_1$}};
\draw[-latex] (\x+7, \y+0.75)++(180:0.25)-- ++(0:0.5);
\draw (\x+7.5, \y+0.5) node {\Large{$A^\prime$}};
\draw[-latex] (\x+8, \y+0.75)++(180:0.25)-- ++(0:0.5);
\draw[thick] (\x+8, \y+1)-- ++(0:1)-- ++(270:2)-- ++(180:1);
}}

\foreach \x in {}{
\foreach \y in {}{
\draw[thick] (\x, \y+1)-- ++(180:1)-- ++(270:2)-- ++(0:1);
\draw[-latex] (\x, \y+0.75)++(0:0.25)-- ++(180:0.5); 
\draw (\x+0.5, \y+0.5) node {\Large{$B$}};
\draw[-latex] (\x+1, \y+0.75)++(0:0.25)-- ++(180:0.5); 
\draw (\x+1.5, \y+0.5) node {\Large{$G_1$}};
\draw[-latex] (\x+2, \y+0.75)++(0:0.25)-- ++(180:0.5); 
\draw (\x+2.5, \y+0.5) node {\Large{$B$}};
\draw[-latex] (\x+3, \y+0.75)++(0:0.25)-- ++(180:0.5); 
\draw (\x+3.5, \y+0.5) node {\Large{$G_1$}};
\draw[-latex] (\x+4, \y+0.75)++(0:0.25)-- ++(180:0.5); 
\draw (\x+4.5, \y+0.5) node {\Large{$B$}};
\draw[-latex] (\x+5, \y+0.75)++(0:0.25)-- ++(180:0.5); 
\draw (\x+5.5, \y+0.5) node {\Large{$G_1$}};
\draw[-latex] (\x+6, \y+0.75)++(0:0.25)-- ++(180:0.5); 
\draw (\x+6.5, \y+0.5) node {\Large{$G_0$}};
\draw[-latex] (\x+7, \y+0.75)++(0:0.25)-- ++(180:0.5); 
\draw (\x+7.5, \y+0.5) node {\Large{$G_1$}};
\draw[-latex] (\x+8, \y+0.75)++(0:0.25)-- ++(180:0.5); 
}}

\foreach \x in {0}{
\foreach \y in {12, 8}{
\draw[-latex] (\x, \y+0.75)++(180:0.25)-- ++(0:0.5);
\draw (\x+0.5, \y+0.5) node {\large{$P_{zig}$}};
\draw[-latex] (\x+1, \y+0.75)++(180:0.25)-- ++(0:0.5);
\draw (\x+1.5, \y+0.5) node {\large{$P_{zig}$}};
\draw[-latex] (\x+2, \y+0.75)++(180:0.25)-- ++(0:0.5);
\draw (\x+2.5, \y+0.5) node{\large{$P_{zig}$}};
\draw[-latex] (\x+3, \y+0.75)++(180:0.25)-- ++(0:0.5);
\draw (\x+3.5, \y+0.5) node {\large{$P_{zig}$}};
\draw[-latex] (\x+4, \y+0.75)++(180:0.25)-- ++(0:0.5);
\draw (\x+4.5, \y+0.5) node {\large{$P_{zig}$}};
\draw[-latex] (\x+5, \y+0.75)++(180:0.25)-- ++(0:0.5);
\draw (\x+5.5, \y+0.5) node {\large{$P_{zig}$}};
\draw[-latex] (\x+6, \y+0.75)++(180:0.25)-- ++(0:0.5);
\draw (\x+6.5, \y+0.5) node {\large{$P_{zig}$}};
\draw[-latex] (\x+7, \y+0.75)++(180:0.25)-- ++(0:0.5);
\draw (\x+7.5, \y+0.5) node {\large{$P_{zig}$}};
\draw[-latex] (\x+8, \y+0.75)++(180:0.25)-- ++(0:0.5);
\draw[thick] (\x+8, \y+1)-- ++(0:1)-- ++(270:2)-- ++(180:1);
}}

\foreach \x in {0}{
\foreach \y in {11, 7}{
\draw[thick] (\x, \y+1)-- ++(180:1)-- ++(270:2)-- ++(0:1)-- ++(90:2);
\draw[-latex] (\x, \y+0.25)++(0:0.25)-- ++(180:0.5); 
\draw (\x+0.5, \y+0.5) node {\large{$P_{zag}$}};
\draw[-latex] (\x+1, \y+0.25)++(0:0.25)-- ++(180:0.5); 
\draw (\x+1.5, \y+0.5) node {\large{$P_{zag}$}};
\draw[-latex] (\x+2, \y+0.25)++(0:0.25)-- ++(180:0.5); 
\draw (\x+2.5, \y+0.5) node {\large{$P_{zag}$}};
\draw[-latex] (\x+3, \y+0.25)++(0:0.25)-- ++(180:0.5); 
\draw (\x+3.5, \y+0.5) node {\large{$P_{zag}$}};
\draw[-latex] (\x+4, \y+0.25)++(0:0.25)-- ++(180:0.5); 
\draw (\x+4.5, \y+0.5) node {\large{$P_{zag}$}};
\draw[-latex] (\x+5, \y+0.25)++(0:0.25)-- ++(180:0.5); 
\draw (\x+5.5, \y+0.5) node {\large{$P_{zag}$}};
\draw[-latex] (\x+6, \y+0.25)++(0:0.25)-- ++(180:0.5); 
\draw (\x+6.5, \y+0.5) node {\large{$P_{zag}$}};
\draw[-latex] (\x+7, \y+0.25)++(0:0.25)-- ++(180:0.5); 
\draw (\x+7.5, \y+0.5) node {\large{$P_{zag}$}};
\draw[-latex] (\x+8, \y+0.25)++(0:0.25)-- ++(180:0.5); 
}}

\foreach \x in {0}{
\foreach \y in {15}{
\draw[-triangle 90, blue, dotted, very thick] (\x+0.5, \y) -- ++(270:11) node [below] {\Large{$N$}};
\draw[-triangle 90, dotted, very thick] (\x+1.5, \y-6.5) -- ++(270:4.5) node [below] {\Large{$1$}};
\draw[red, dotted, very thick] (\x+2.5, \y) -- ++(270:5.5);
\draw[-triangle 90, blue, dotted, very thick] (\x+2.5, \y-5.5) -- ++(270:5.5) node [below] {\Large{$N$}};
\draw[-triangle 90, dotted, very thick] (\x+3.5, \y-6.5) -- ++(270:4.5) node [below] {\Large{$0$}};
\draw[-triangle 90, blue, dotted, very thick] (\x+4.5, \y) -- ++(270:11) node [below] {\Large{$N$}};
\draw[-triangle 90, dotted, very thick] (\x+5.5, \y-6.5) -- ++(270:4.5) node [below] {\Large{$0$}};
\draw[-triangle 90, blue, dotted, very thick] (\x+6.5, \y) -- ++(270:11) node [below] {\Large{$N$}};
\draw[-triangle 90, dotted, very thick] (\x+7.5, \y-6.5) -- ++(270:4.5) node [below] {\Large{$0$}};
}}

\end{tikzpicture}}

\caption{Example run of Module 4 (due to the space shortage, the last 3 zigzags are omitted). 
Here the transition $f_2$ has been chosen so that only the corresponding $(2{\times}2{-}1)$-th zig ends at the bottom. 
As a result, only the 3rd zag outputs the hardcoded target state 1000 below. 
All the succeeding zigzags propagate 1000 downward. 
}
\label{fig:example_module4}
\end{figure}

\paragraph{Module 4 (outputting the target state of the transition chosen)} folds into $2n$ zigzags. 
Its $(2k{-}1)$-th zig checks whether $f_k$ was chosen or not, and if it was, the next zag sets $z_{q_i[j]}$ to $t_k[j]$ (recall $t_k$ is the target of $f_k$). 

The transcript for the $(2k{-}1)$-th zig is represented semantically as 
\begin{equation}\label{transcript:m4-zig}
(A'A')^{k-1} A_1A' (A'A')^{n-k}. 
\end{equation}
Observe that the sole $A_1$ is positioned so as to read the 1-bit of whether $f_k$ was chosen or not. 
The zig starts at the bottom. 
Since $A'$ always start and end at the same height (Figure~\ref{fig:brick_A'}), the $A_1$ starts at the bottom. 
It ends at the bottom if it reads $Y$, or top otherwise (Figure~\ref{fig:brick_A1}). 
The succeeding turner is functional, which lets the next zag start at the bottom if the previous zig has ended at the bottom, or at the top otherwise. 
In this way, the $(2k{-}1)$-th zag starts at the bottom iff Module 3 has chosen $f_k$. 

\begin{figure}[tb]
\centering
\includegraphics[width=0.8\linewidth]{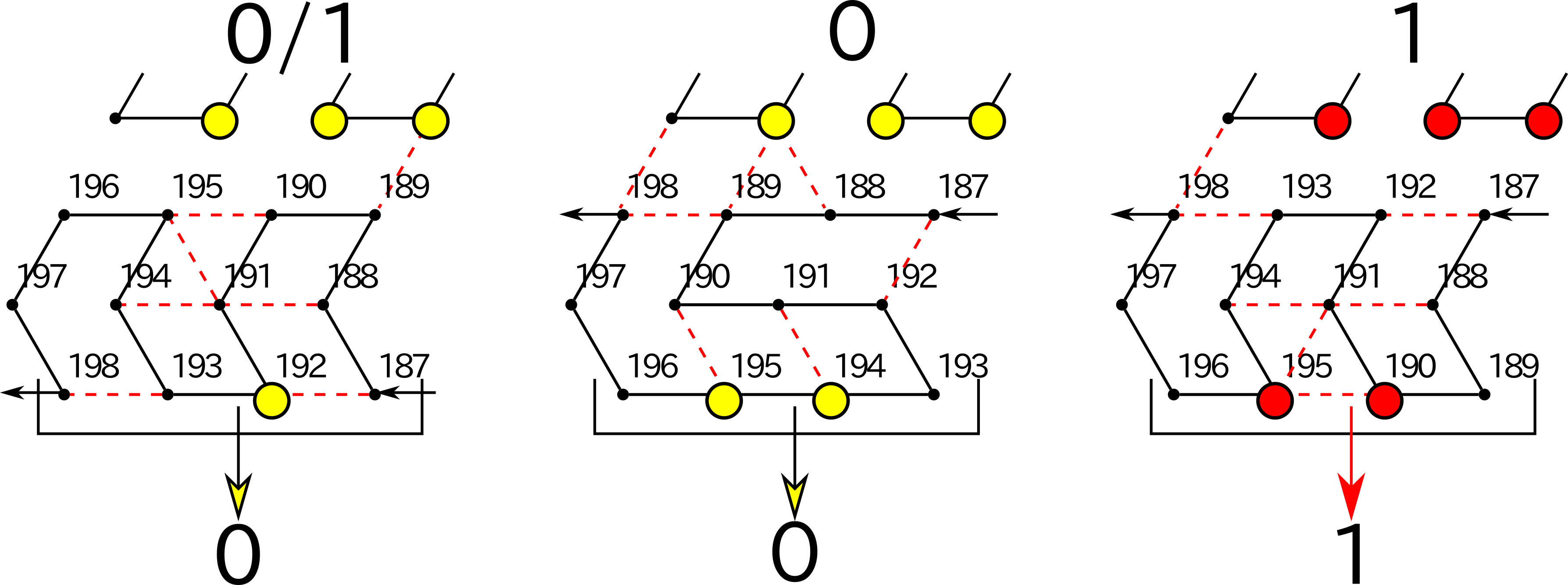}
\caption{The three bricks of $G_0$, that is, $G_{\rm 0-\ast b}$, $G_{\rm 0-0t}$, and $G_{\rm 0-1t}$.}
\label{fig:brick_G0}
\end{figure}

\begin{figure}[tb]
\centering
\includegraphics[width=0.8\linewidth]{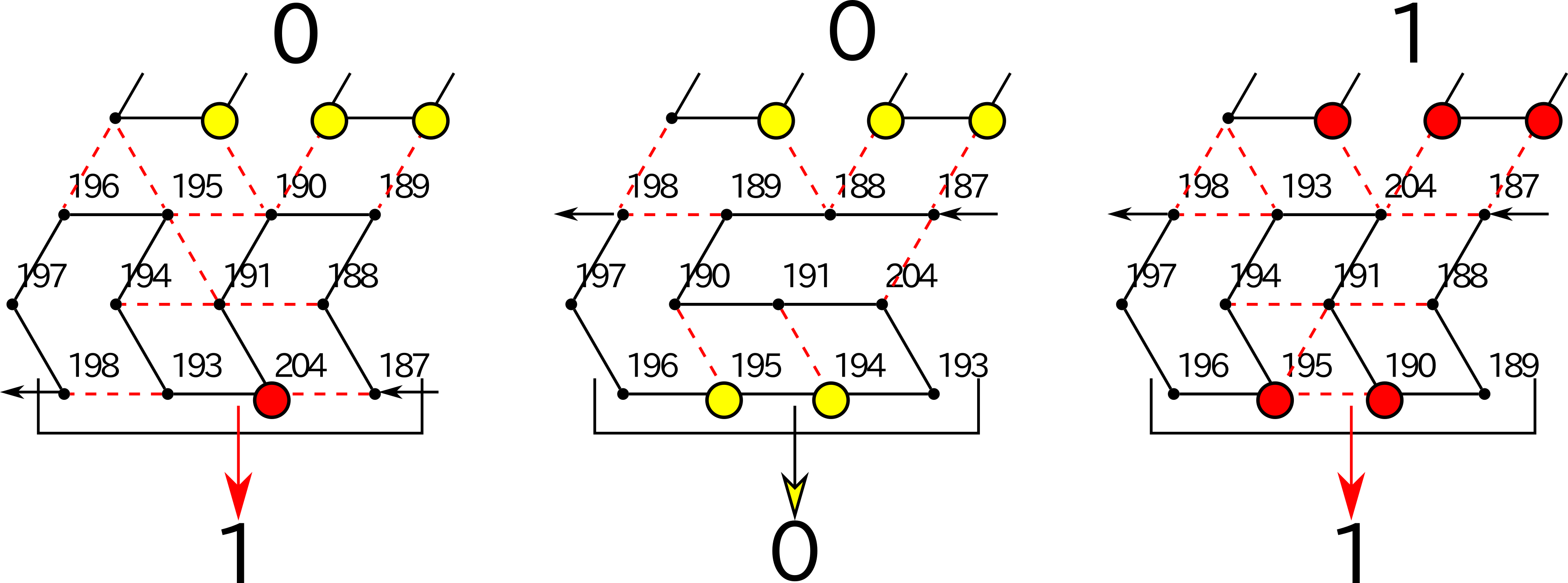}
\caption{The three bricks of $G_1$, that is, $G_{\rm 1-0b}$, $G_{\rm 1-0t}$, and $G_{\rm 1-1t}$.}
\label{fig:brick_G1}
\end{figure}

The transcript for the $(2k{-}1)$-th zag is represented semantically as 
\begin{equation}\label{transcript:m4-zag}
	\bigl(\mbox{$\bigodot_{j=n}^{k+1}$} (G_{t_k[j]} B) \bigr) G_{t_k[k]} G_0 \bigl( \mbox{$\bigodot_{j=k-1}^{1}$} (G_{t_k[j]} B) \bigr).
\end{equation}
All the bricks of submodules $G_0$ and $G_1$ (see Figures~\ref{fig:brick_G0} and \ref{fig:brick_G1}) start and end at the same height; thus propagating the 1-bit of whether $f_k$ was chosen or not (bottom means chosen) through this zag. 
Note that this transcript is transcribed from right to left so that these $G_0$'s and $G_1$'s read $z$-variables. 
$G_0$ and $G_1$ just copy what they read downward if they start at the top, that is, in all zags but the one corresponding to the chosen transition. 
In the ``chosen'' zag, they rather output 0 and 1 downward, respectively. 
Comparing \eqref{transcript:m4-zig} with \eqref{transcript:m4-zag}, we can observe that below the sole instance of $A_0$ is transcribed an instance of $G_0$. 
This $G_0$ plays a different role from other $G_0$'s in the zag. 
The $A_1$ above outputs $Y$ or $N$ depending on whether $f_k$ was chosen or not. 
If it outputs $Y$, then the $(2k{-}1)$-th zag starts at the bottom as just mentioned, and the sole brick of $G_0$ that starts at the bottom outputs $0 = N$. 
Otherwise, $G_0$ just propagates its output $0 = N$ downward. 
In this way, all the $x$-variables are initialized to $N$ for the sake of succeeding period. 

	\subsection{Verification}

Using a simulator developed for \cite{MasudaSekiUbukata2018}, we have checked for each submodule that it folds as expected in all the expected environments. 
An expected environment can be described in terms of bricks of surrounding submodules. 
Folding of a submodule into a brick in an environment causes a transition to another environment for the next submodule. 
Such transitions combine all the expected environments together into one closed system called brick automaton, whose vertices are expected environments described in terms of surrounding bricks. 

The automaton is so large that it could not help but be split into parts for presentation. 
Spacers are depicted as a cyan rectangle. 
Transitions are labeled with {\tt T} or {\tt B}, which means that the previous brick has ended folding at the top or bottom. 

\begin{figure}[h]
\centering
\includegraphics[width=\linewidth]{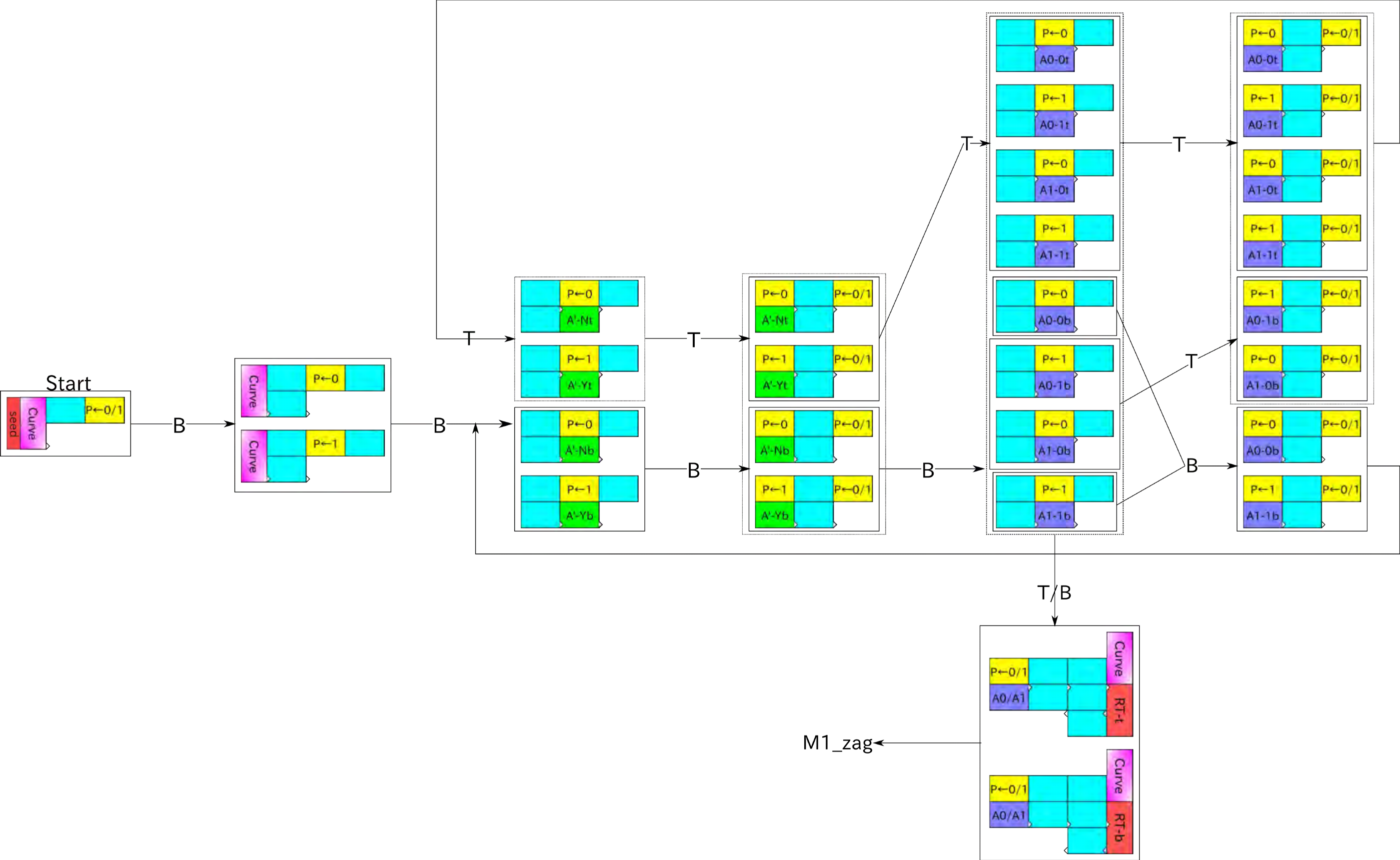}
\caption{The part {\tt M1\_zig} of the brick automaton for the $(2k{-}1)$-th zig of Module 1, where yellow rectangles represent $P_{\rm zag}$ in the previous formatting zag.}
\label{ba:module1_zig}
\end{figure}

\begin{figure}[p]
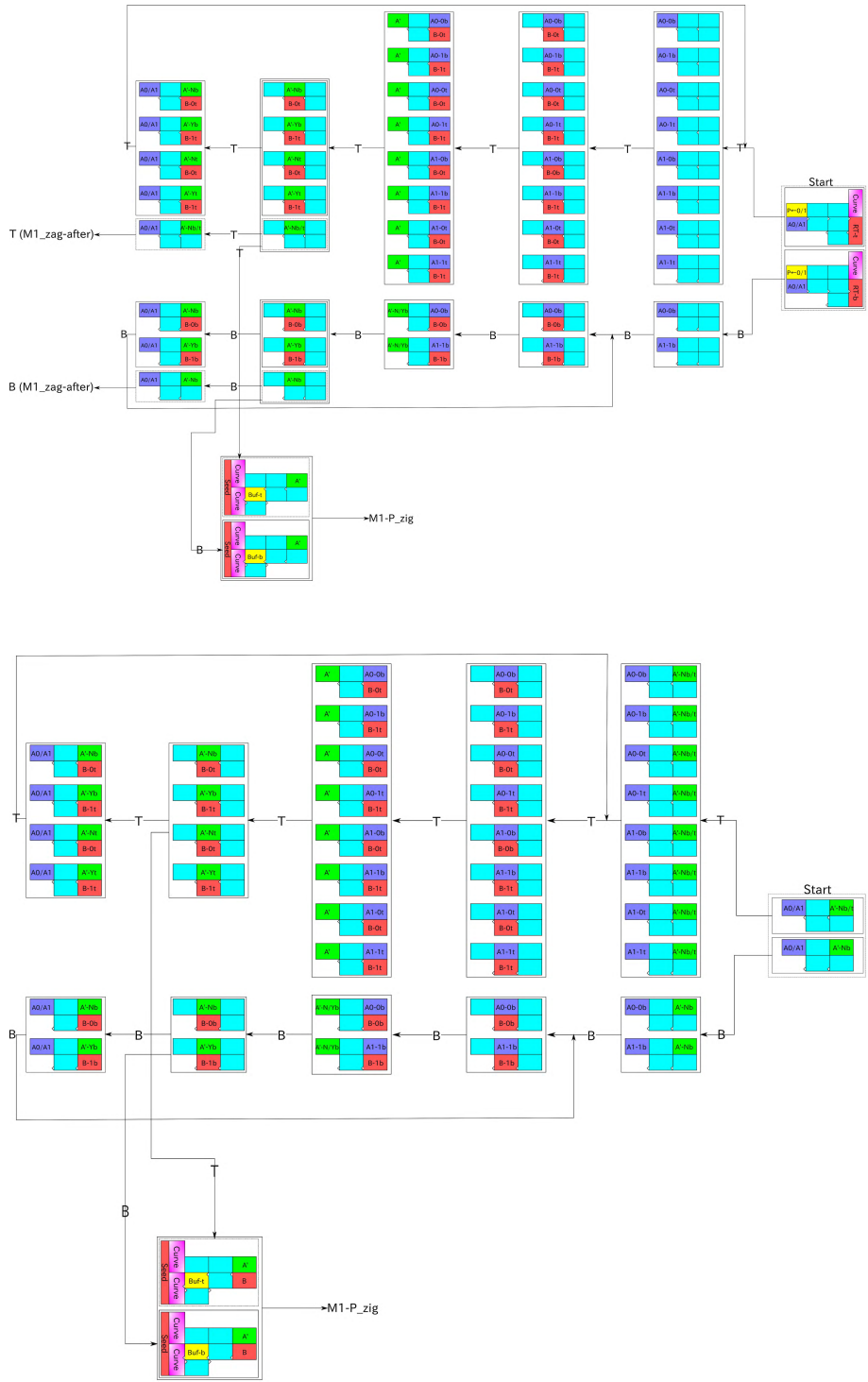

\centering
\includegraphics[width=\linewidth]{M1_zag-before.pdf}
\vspace*{5mm}

\includegraphics[width=\linewidth]{M1_zag-after.pdf}
\caption{Two parts of the brick automaton for the $(2k{-}1)$-th zag of Module 1 (Top) {\tt M1\_zag-before}, which describes transitions among bricks for this zag until the zag encounters the g-spacer $B'$ (depicted as a cyan rectangle like the other spacers), and (Bottom) {\tt M1\_zag-after} for describing transitions after the encounter.}
\label{ba:module1_zag}
\end{figure}

\begin{figure}[p]
\centering
\includegraphics[width=\linewidth]{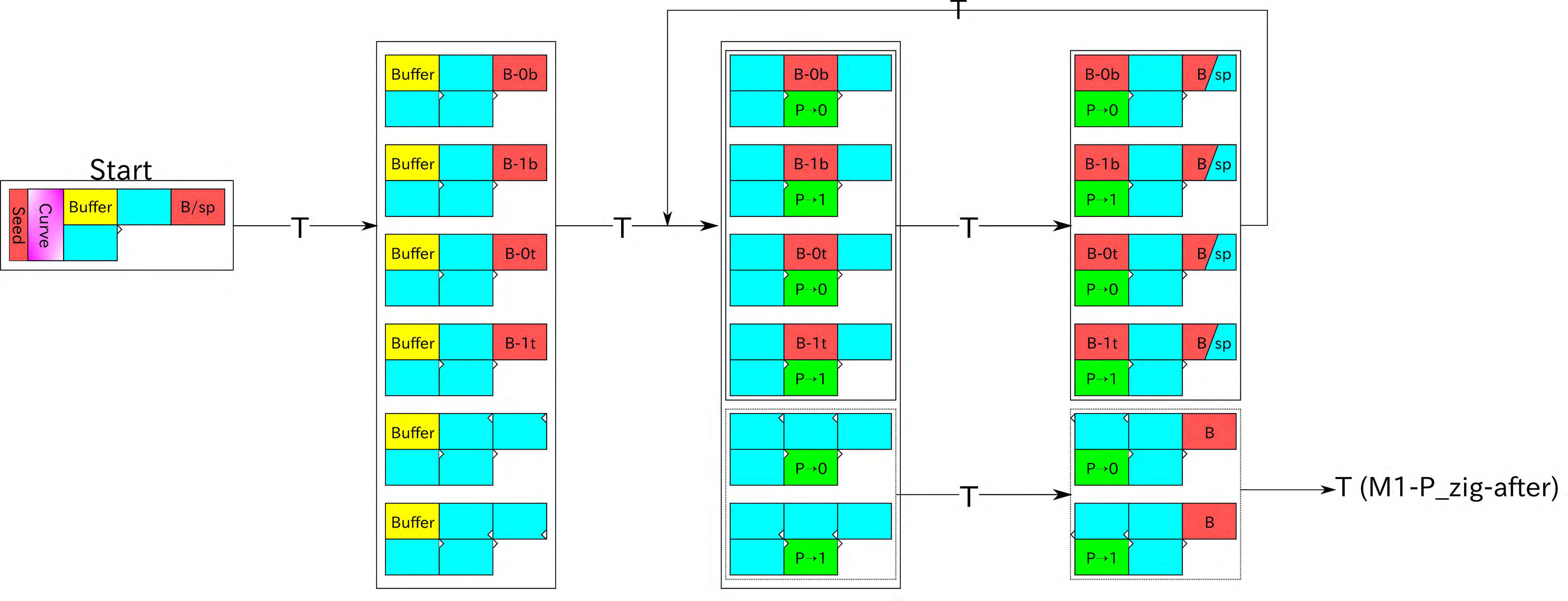}
\vspace*{5mm}

\includegraphics[width=0.8\linewidth]{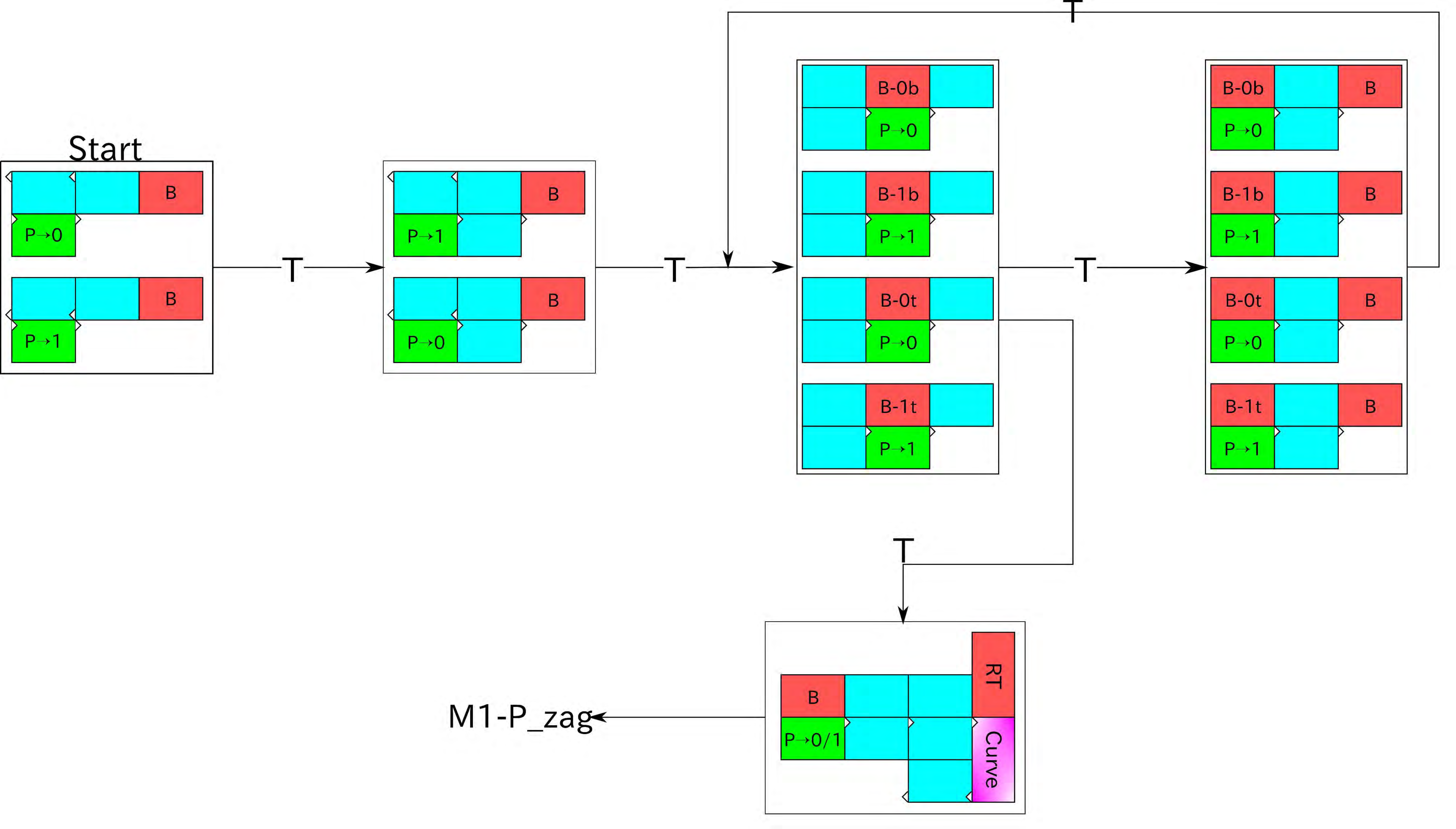}
\vspace*{5mm}

\includegraphics[width=\linewidth]{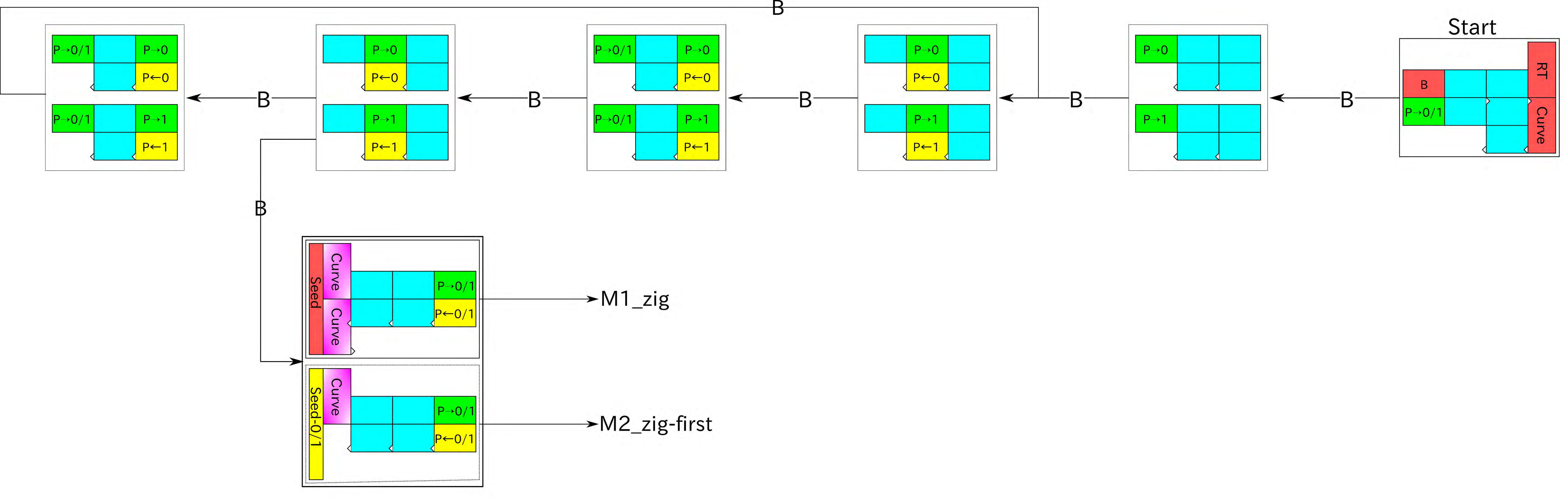}
\caption{Three parts of the brick automaton: (Top, Middle) {\tt M1-P\_zig} and {\tt M1-P\_zig-after} for the $2k$-th (formatting) zig until $B'$ and after $B'$, respectively,  and (Bottom) {\tt M1-P\_zag} for the $2k$-th zag of Module 1.}
\label{ba:module1_formatting}
\end{figure}

\paragraph{Module 1.} 
The parts of the brick automaton for Module 1 are illustrated in Figures~\ref{ba:module1_zig}, \ref{ba:module1_zag}, and \ref{ba:module1_formatting}. 
Recall that the $(2k{-}1)$-th zag involves exactly one instance of $B'$. 
The parts for the $(2k{-}1)$-th zag and the succeeding $2k$-th zig therefore get so large that they are split into two, respectively in Figure~\ref{ba:module1_zag} and Figure~\ref{ba:module1_formatting} (Top, Middle). 

\begin{figure}[p]
\centering
\includegraphics[width=\linewidth]{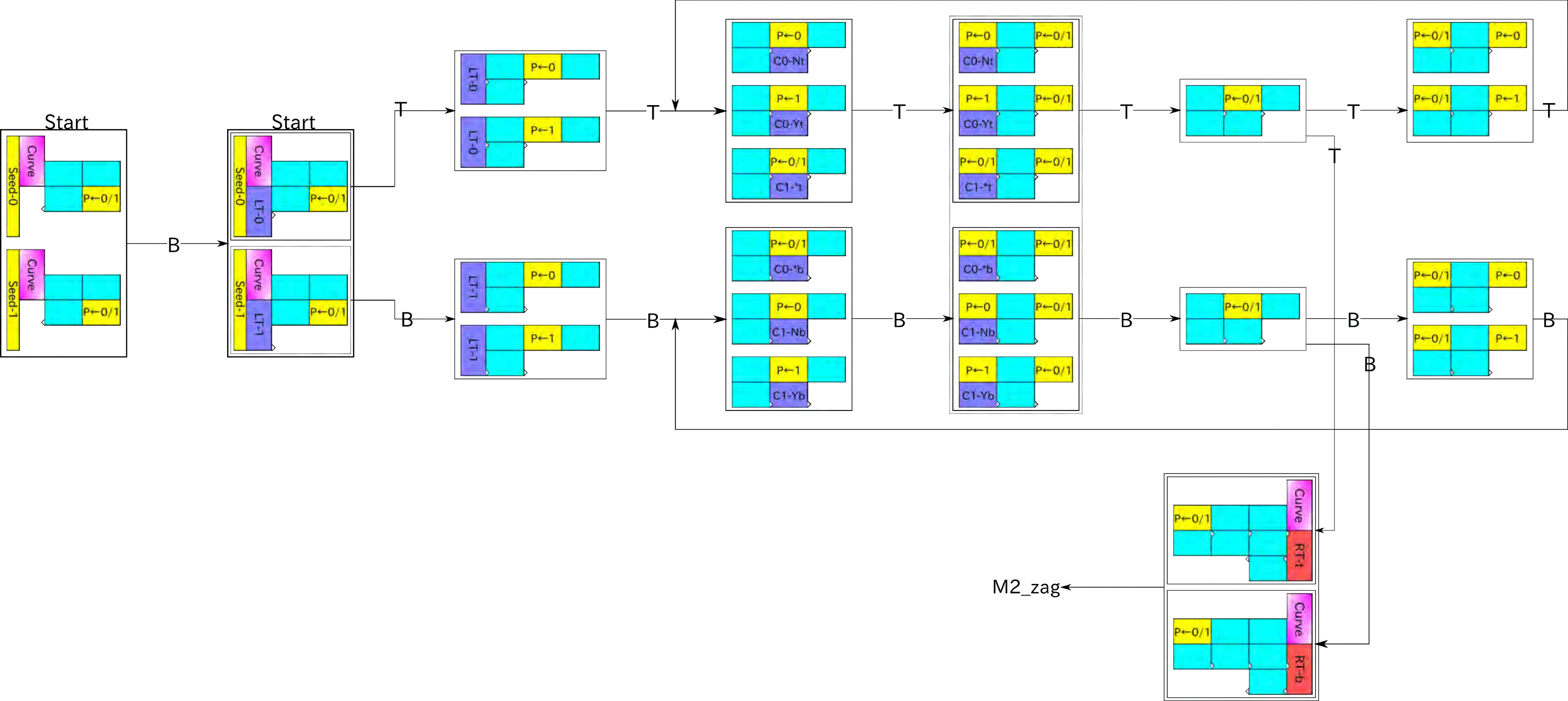}
\vspace*{5mm}

\includegraphics[width=\linewidth]{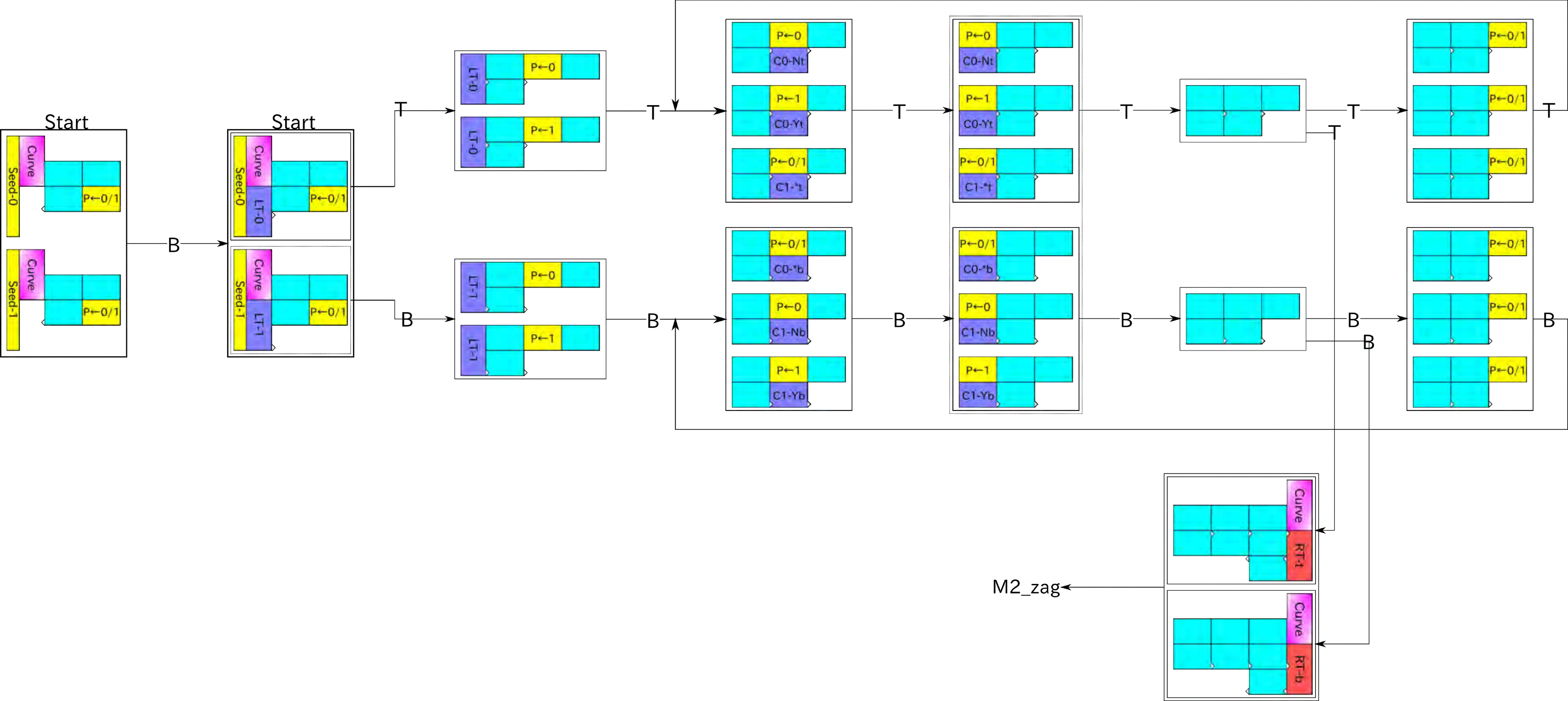}
\vspace*{5mm}

\includegraphics[width=0.8\linewidth]{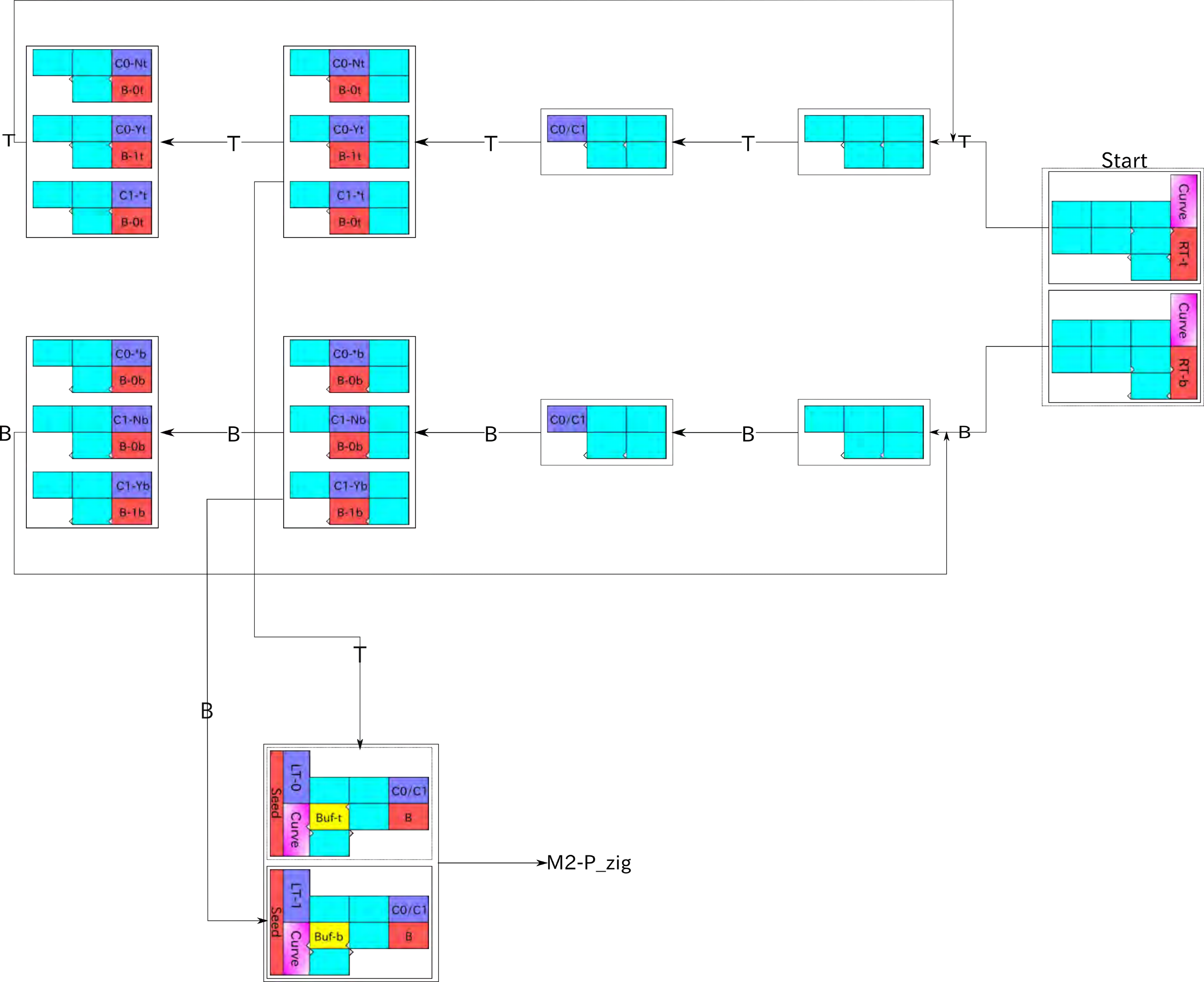}
\caption{Three parts of the brick automaton: (Top) {\tt M2\_zig-first} for the 1st zig, (Middle) {\tt M2\_zig} for the $(2k{-}1)$-th zig for $k \ge 2$, and (Bottom) {\tt M2\_zag} for the $(2k{-}1)$-th zag of Module 2 for $k \ge 1$.}
\label{ba:M2_zigzag}
\end{figure}

\begin{figure}[p]
\centering
\includegraphics[width=\linewidth]{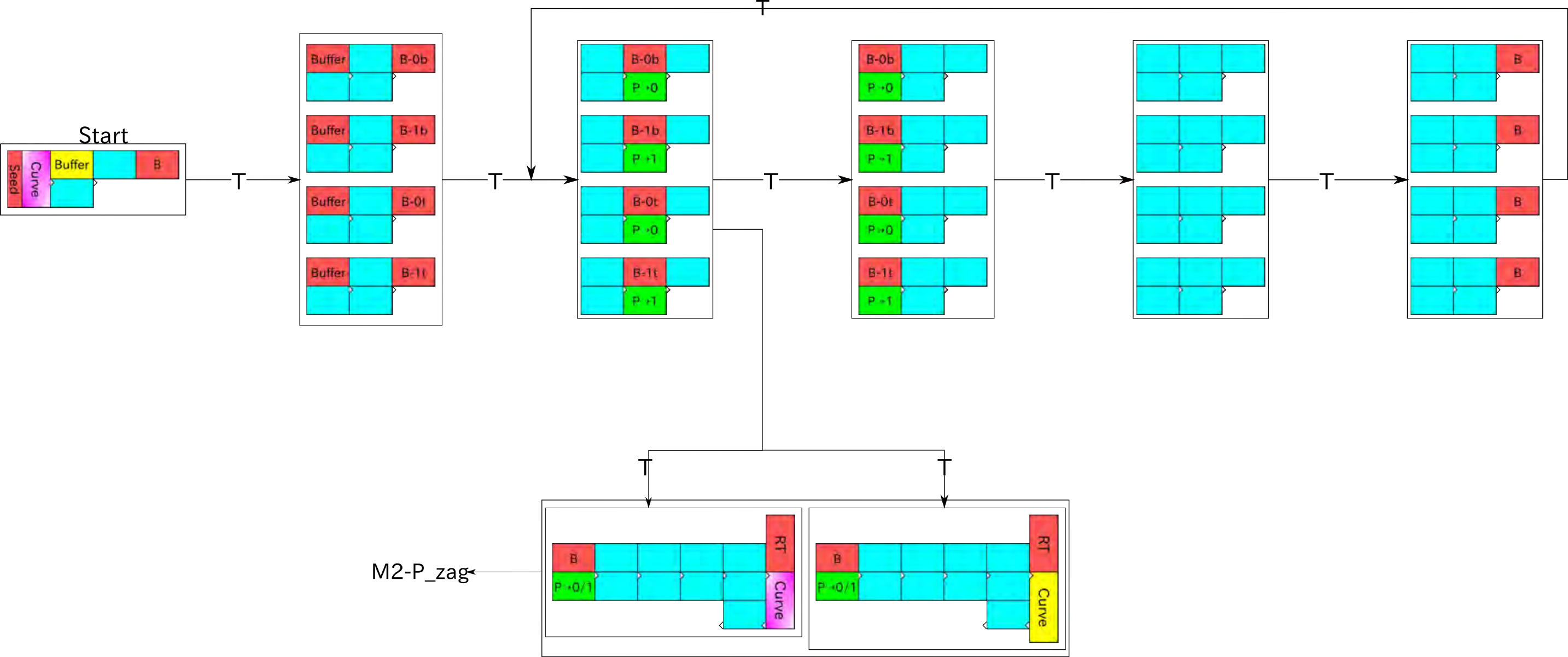}

\vspace*{5mm} 

\includegraphics[width=\linewidth]{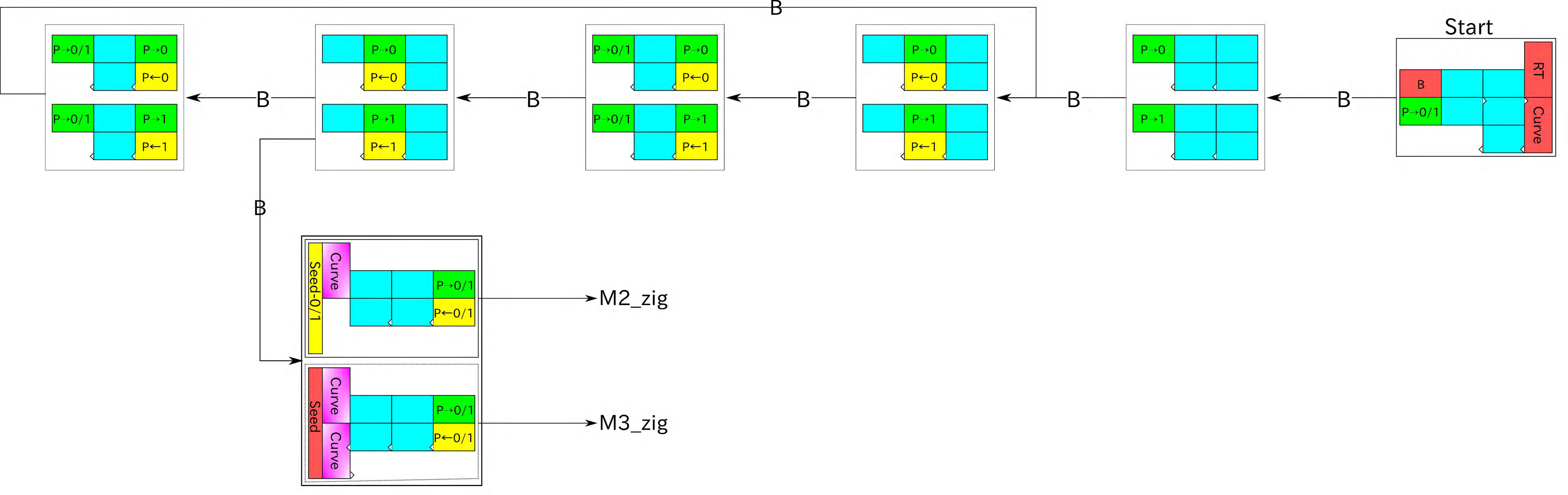}
\caption{Two parts of the brick automaton: (Top) {\tt M2-P\_zig} for the $2k$-th (formatting) zig and (Bottom) {\tt M2-P\_zag} for the $2k$-th (formatting) zag of Module~2.}
\label{ba:M2_formatting}
\end{figure}

\paragraph{Module 2.} 
The parts of the brick automaton for Module 2 are illustrated in Figures~\ref{ba:M2_zigzag} and \ref{ba:M2_formatting}. 
As mentioned previously, this module does not have to propagate the $n$-bits to identify the current state $q_{n-1}$. 
Thus, in the formatting zigs and zags, the corresponding $n$ instances of $P_{\rm zig}$ and of $P_{\rm zag}$ are in fact replaced by a glider. 
As a result, the 1st zig and the other $(2k{-}1)$-th zigs (for $k \ge 2$) encounter different environments. 
This difference is illustrated in Figure~\ref{ba:M2_zigzag} (Top, Middle). 

\begin{figure}[p]
\centering
\includegraphics[width=\linewidth]{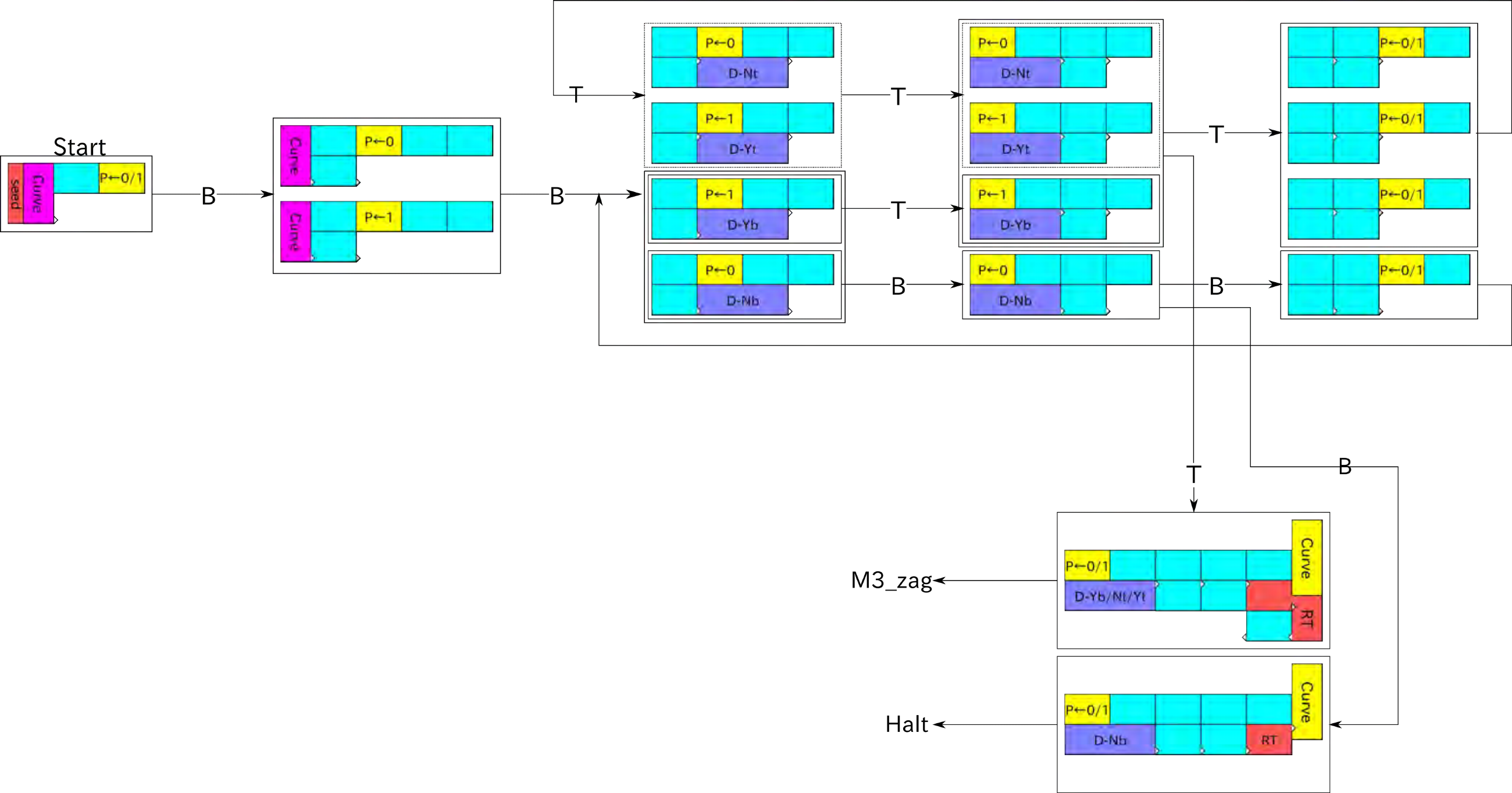}
\vspace*{5mm}

\includegraphics[width=\linewidth]{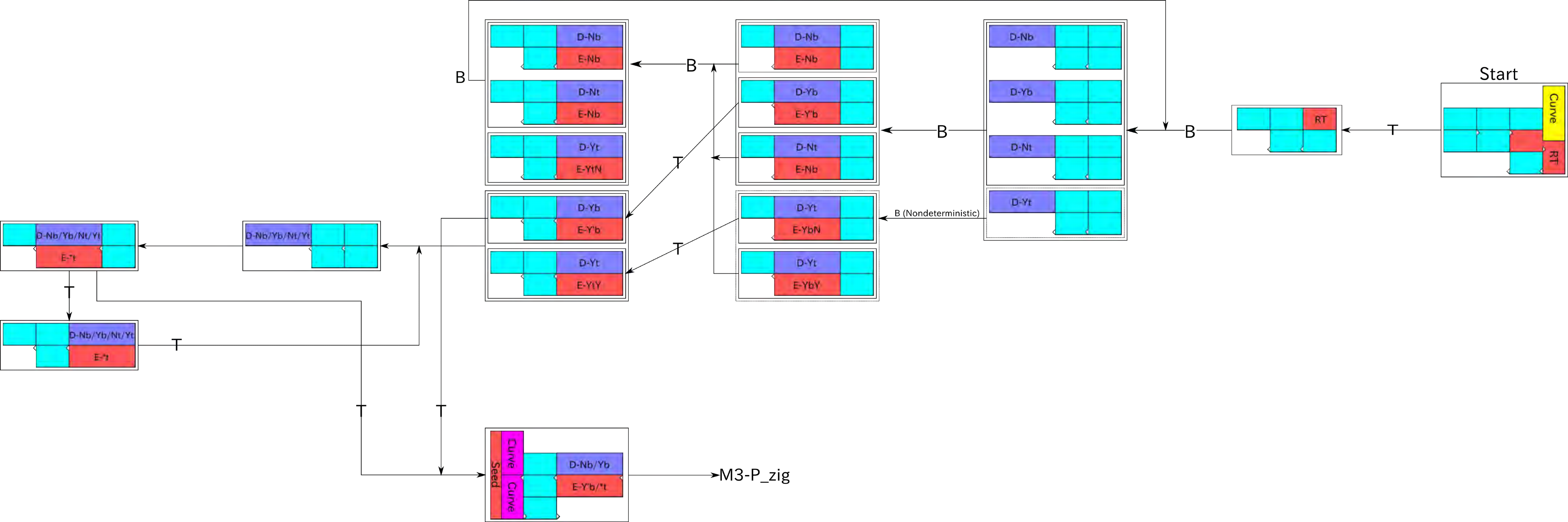}
\caption{Two parts of the brick automaton: (Top) {\tt M3\_zig} for the 1st zig and (Bottom) {\tt M3\_zag} for the 1st zag of Module 3.}
\label{ba:M3_zigzag}
\end{figure}

\begin{figure}[p]
\centering
\includegraphics[width=\linewidth]{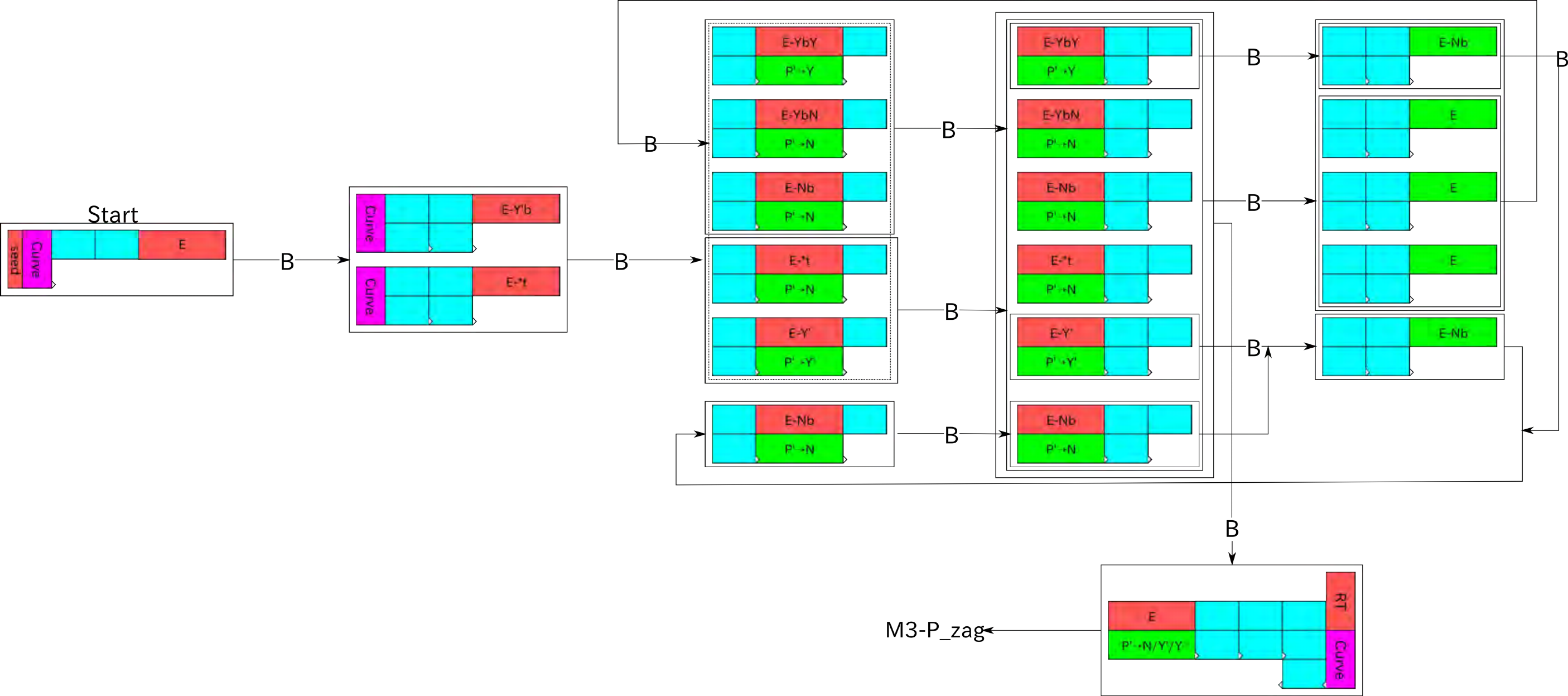}
\vspace*{5mm}

\includegraphics[width=\linewidth]{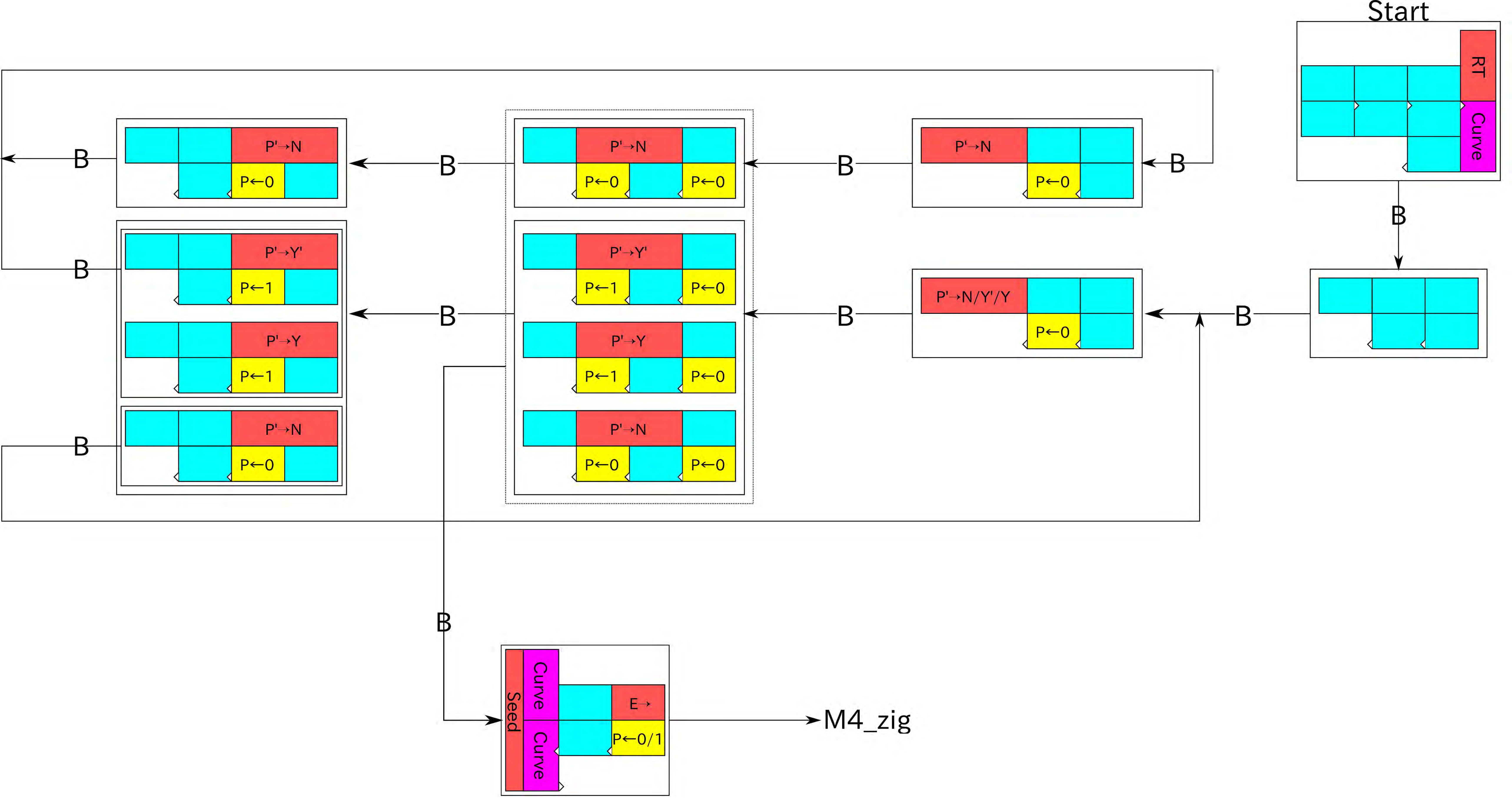}
\caption{Two parts of the brick automaton: (Top) {\tt M3-P\_zig} for the 2nd (formatting) zig and (Bottom) {\tt M3-P\_zag} for the 2nd zag of Module~3.}
\label{ba:M3_formatting}
\end{figure}

\paragraph{Module 3.} 
The parts of the brick automaton for Module 3 are illustrated in Figures~\ref{ba:M3_zigzag} and \ref{ba:M3_formatting}.

\begin{figure}[h]
\centering
\includegraphics[width=\linewidth]{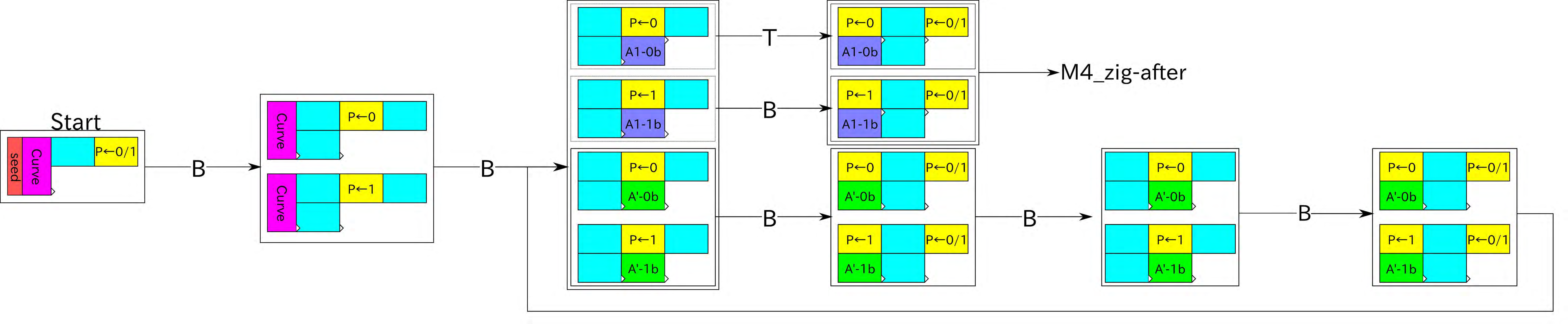}
\vspace*{5mm}

\includegraphics[width=0.7\linewidth]{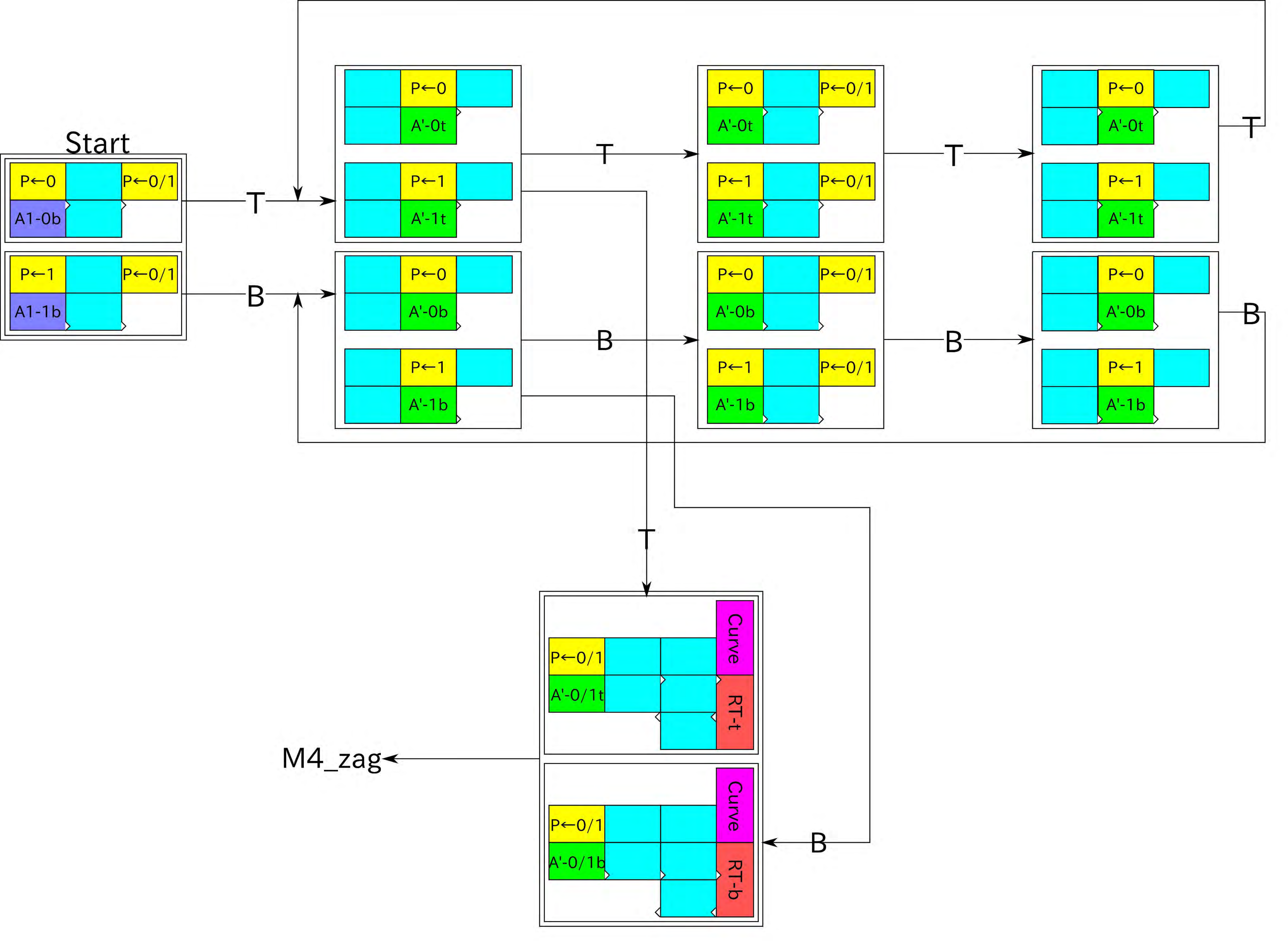}
\caption{Two parts of the brick automaton: (Top) {\tt M4\_zig-before}, which describes transitions until the sole instance of $A_1$ is transcribed, and (Bottom) {\tt M4\_zig-after} for the transitions after that $A_1$. 
}
\label{ba:M4_zig}
\end{figure}

\begin{figure}[p]
\centering
\includegraphics[width=\linewidth]{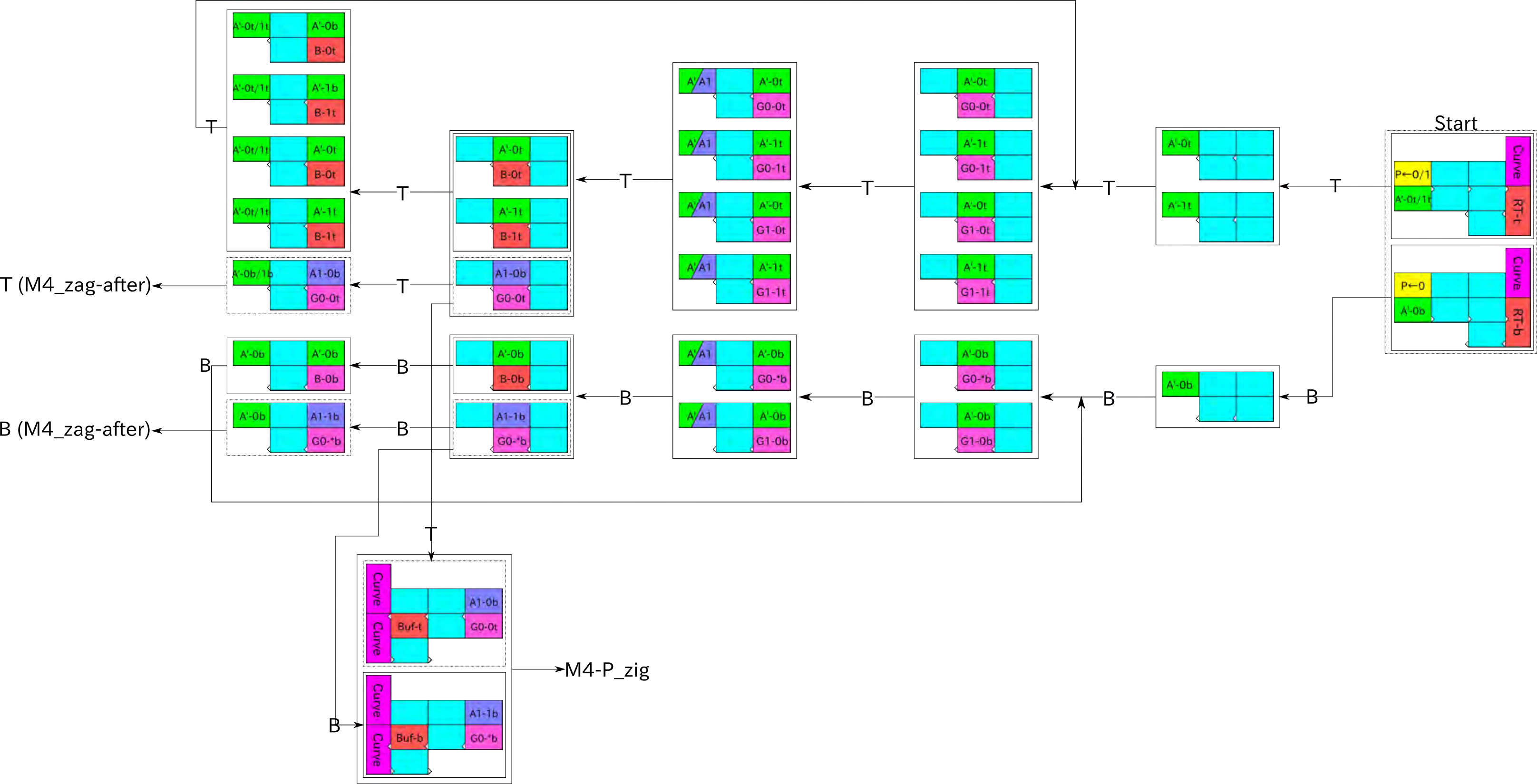}
\vspace*{5mm}

\includegraphics[width=\linewidth]{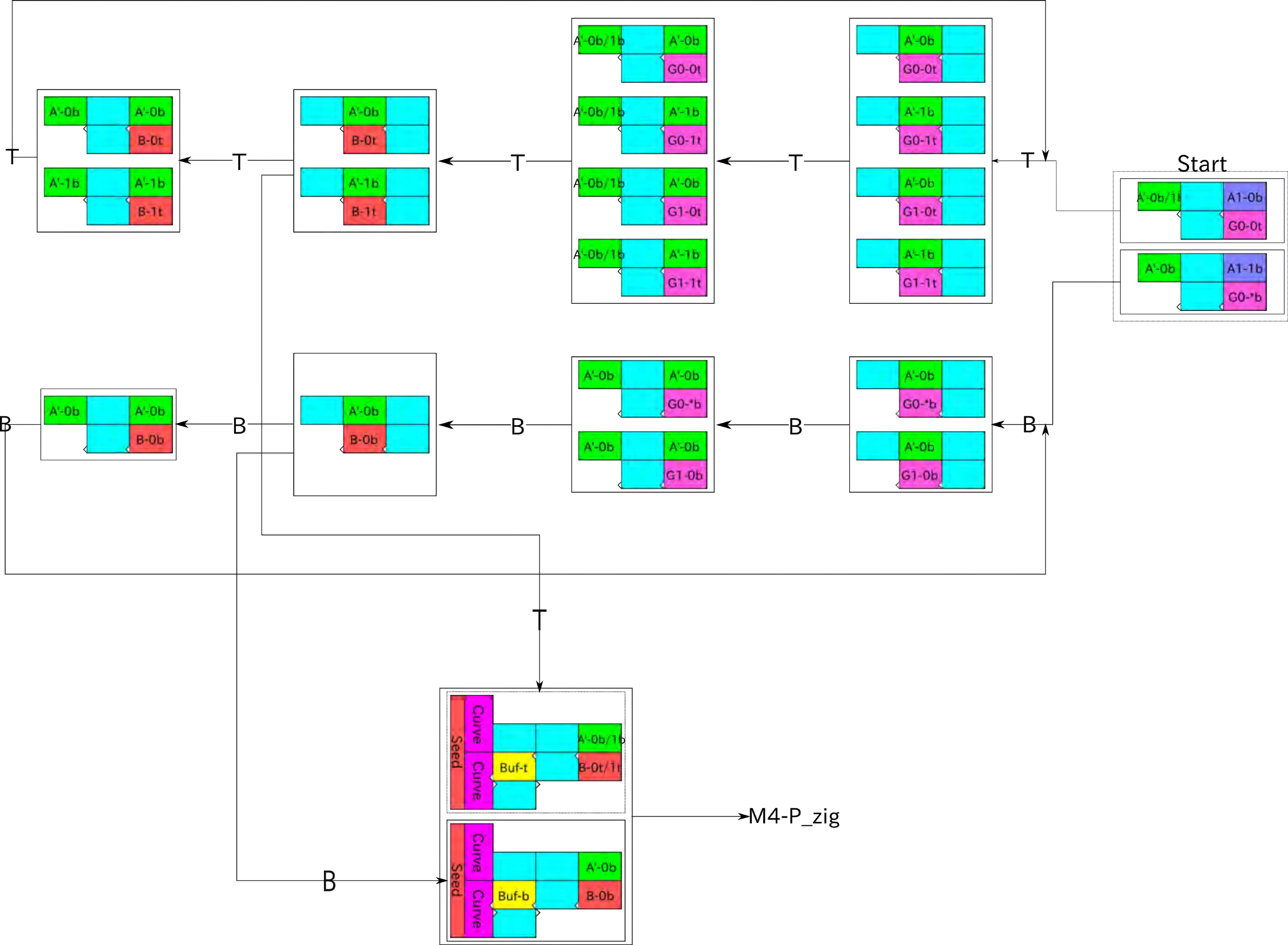}
\caption{Two parts {\tt M4\_zag} of the brick automaton for the $(2k{-}1)$-th zag of Module 4: 
(Top) {\tt M4\_zag-before}, which describes transitions until the sole instance of $A_1$ in the previous zig is encountered, and (Bottom) {\tt M4\_zag-after} for the transitions after the encounter. 
}
\label{ba:M4_zag}
\end{figure}

\begin{figure}[h]
\centering
\includegraphics[width=\linewidth]{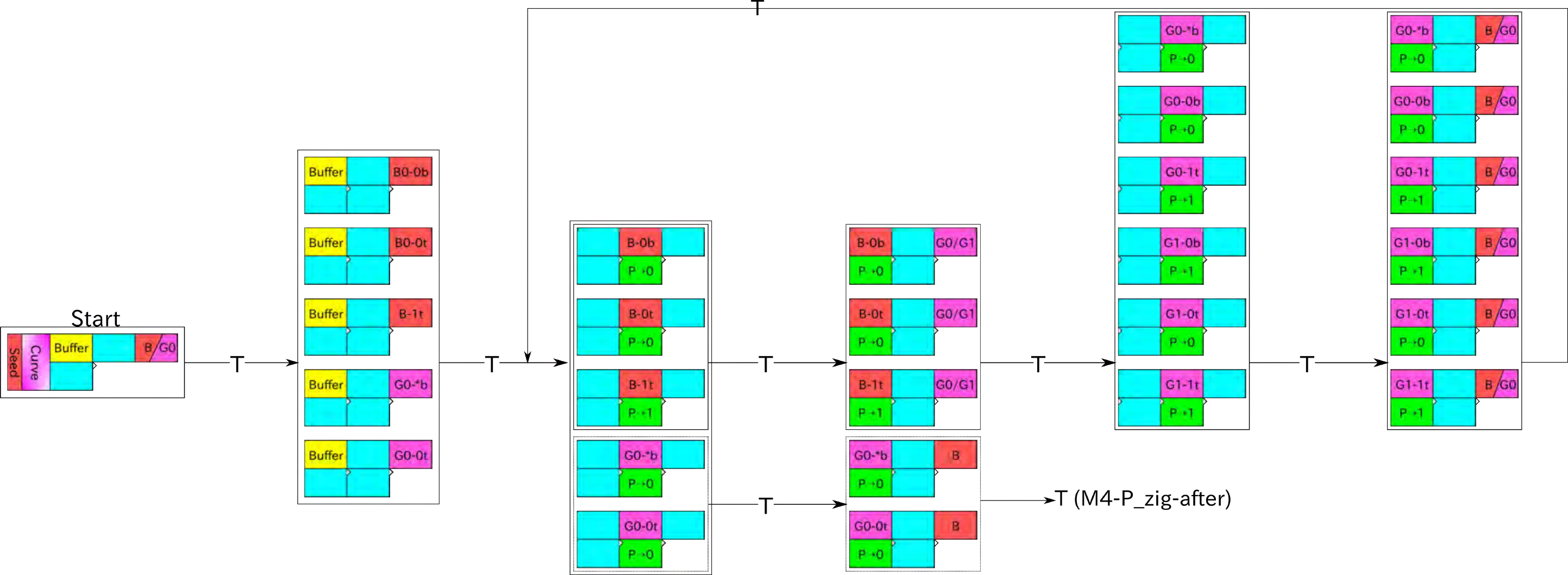}
\vspace*{5mm}

\includegraphics[width=\linewidth]{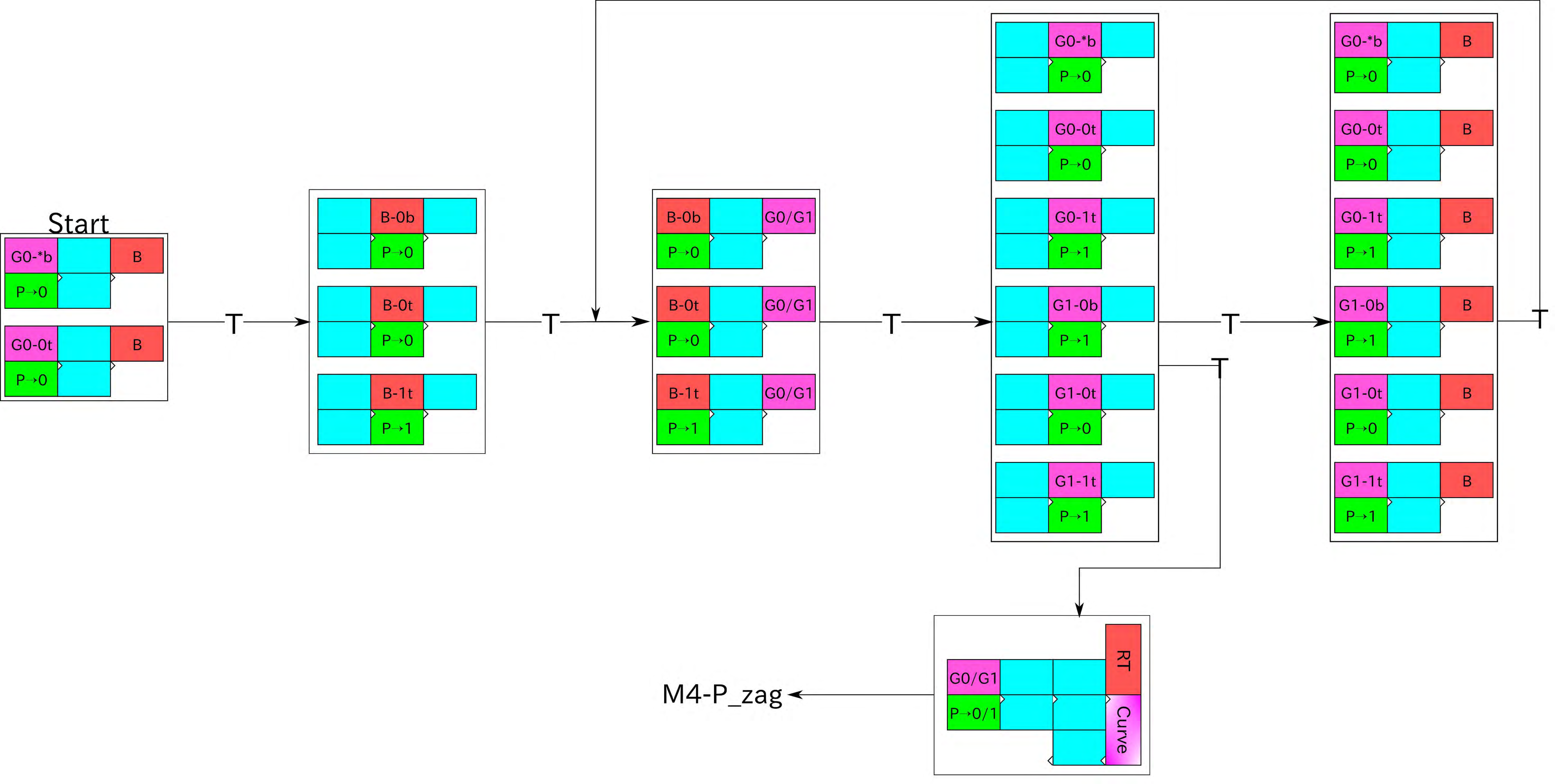}
\caption{Two parts of the brick automaton: (Top) {\tt M4-P\_zig-before} for the $2k$-th (formatting) zig of Module 4 until the zig encounters the sole instance of $A_1$, and (Bottom) {\tt M4-P\_zig-after} after of $A_1$.}
\label{ba:M4_formatting_zig}
\end{figure}

\begin{figure}[h]
\centering
\includegraphics[width=\linewidth]{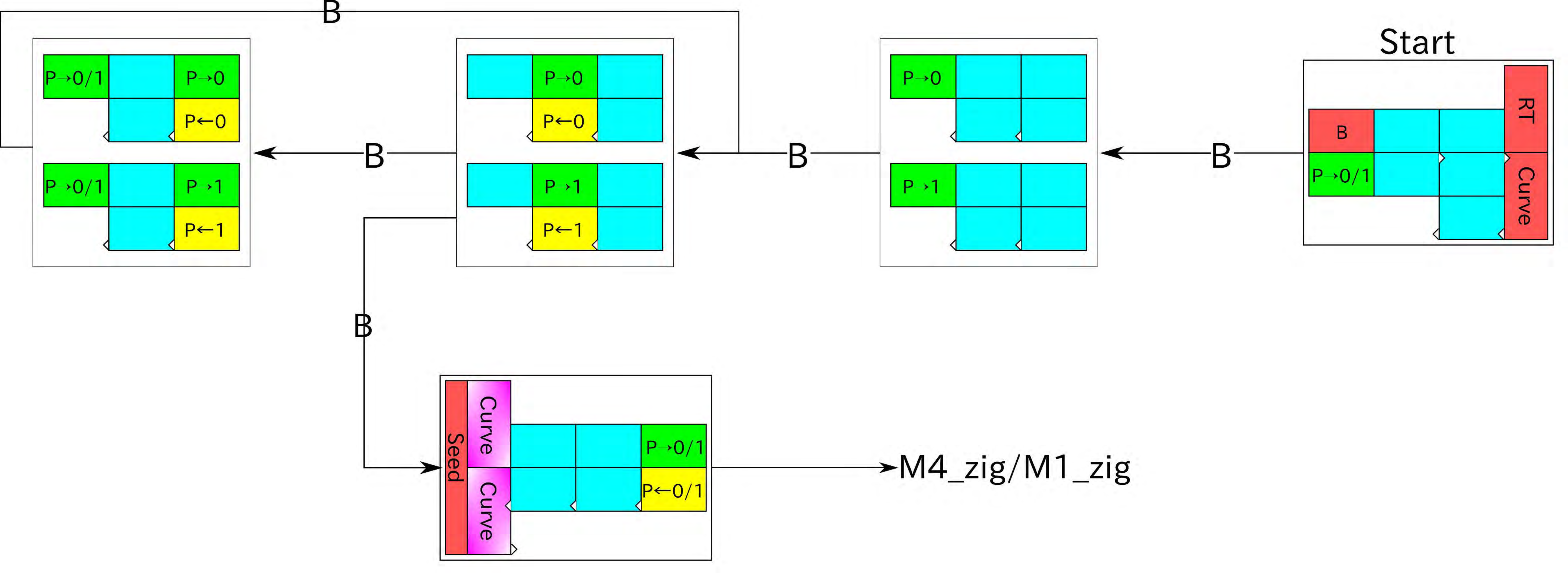}
\caption{Two part {\tt M4-P\_zag} of the brick automaton for the $2k$-th (formatting) zag of Module 4.}
\label{ba:M4_formatting_zag}
\end{figure}

\paragraph{Module 4.} 
The parts of the brick automaton for Module 4 are illustrated in Figures~\ref{ba:M4_zig}, \ref{ba:M4_zag}, \ref{ba:M4_formatting_zig}, and \ref{ba:M4_formatting_zag}. 

	\bibliographystyle{plain}
	\bibliography{os_afas}

\end{document}